%% file: main.tex
\documentclass[sigconf]{acmart}

\copyrightyear{2022} 
\acmYear{2022} 
\setcopyright{acmcopyright}\acmConference[ASE '22]{37th IEEE/ACM International Conference on Automated Software Engineering}{October 10--14, 2022}{Rochester, MI, USA}
\acmBooktitle{37th IEEE/ACM International Conference on Automated Software Engineering (ASE '22), October 10--14, 2022, Rochester, MI, USA}
\acmPrice{15.00}
\acmDOI{10.1145/3551349.3556939}
\acmISBN{978-1-4503-9475-8/22/10}

\acmConference[ASE 2022]{The 37th IEEE/ACM International Conference on Automated Software Engineering}{10-14 October, 2022}{Ann Arbor}

\usepackage{balance}
\usepackage{microtype}
\usepackage{subcaption}
\usepackage{xspace}
\usepackage{xcolor}
\usepackage{amsmath}
\usepackage{hyperref}
\usepackage[capitalise]{cleveref}
\usepackage{booktabs}
\usepackage{comment}
\usepackage{array}
\usepackage{placeins}
\usepackage[ruled,linesnumbered]{algorithm2e}

\include{macros}

\newcommand\definetool[2]{\newcommand{#1}{{\textsc{#2}}\xspace}}
\newcommand{\atwelve}{\ensuremath{\hat{A}_{12}}\xspace}

\definetool{\Scratch}{Scratch}
\definetool{\Whisker}{Whisker}
\definetool{\WhiskerWeb}{WhiskerWeb}
\definetool{\Neatest}{Neatest}
\definetool{\ENEAT}{\Neatest}

\def\titleset{\leftskip0pt\rightskip0pt}
\begin{document}

\title{\titleset Neuroevolution-Based Generation of Tests and Oracles for Games}

\author{Patric Feldmeier}
\affiliation{%
  \institution{University of Passau}
  \country{Germany}}

\author{Gordon Fraser}
\affiliation{%
  \institution{University of Passau}
  \country{Germany}}

\begin{abstract}
	Game-like programs have become increasingly popular in many software engineering domains such as
mobile apps, web applications, or programming education.
	However, creating tests for programs that have the purpose of challenging
human players is a daunting task for automatic test generators. Even if test
generation succeeds in finding a relevant sequence of events to exercise a
program, the randomized nature of games means that it may neither be possible
to reproduce the exact program behavior underlying this sequence, nor to
create test assertions checking if observed randomized game behavior is correct.
	To overcome these problems, we propose \Neatest, a novel test generator
based on the \emph{NeuroEvolution of Augmenting Topologies} (NEAT) algorithm.
\Neatest systematically explores a program's statements, and creates neural
networks that operate the program in order to reliably reach each
statement---that is, \Neatest learns to play the game in a way to reliably
cover different parts of the code. As the networks learn the actual game
behavior, they can also serve as test oracles by evaluating how surprising the
observed behavior of a program under test is compared to a supposedly correct version of the program.
	We evaluate this approach in the context of \Scratch, an educational
programming environment.
	Our empirical study on 25 non-trivial \Scratch games demonstrates that our
approach can successfully train neural networks that are not only far more
resilient to random influences than traditional test suites consisting of
static input sequences, but are also highly effective with an average
mutation score of more than 65\%.
\end{abstract}



\maketitle

\section{Introduction}

\begin{figure}[!tbp]
  \begin{subfigure}[b]{\columnwidth}
    \centering
    \includegraphics[width=.75\columnwidth]{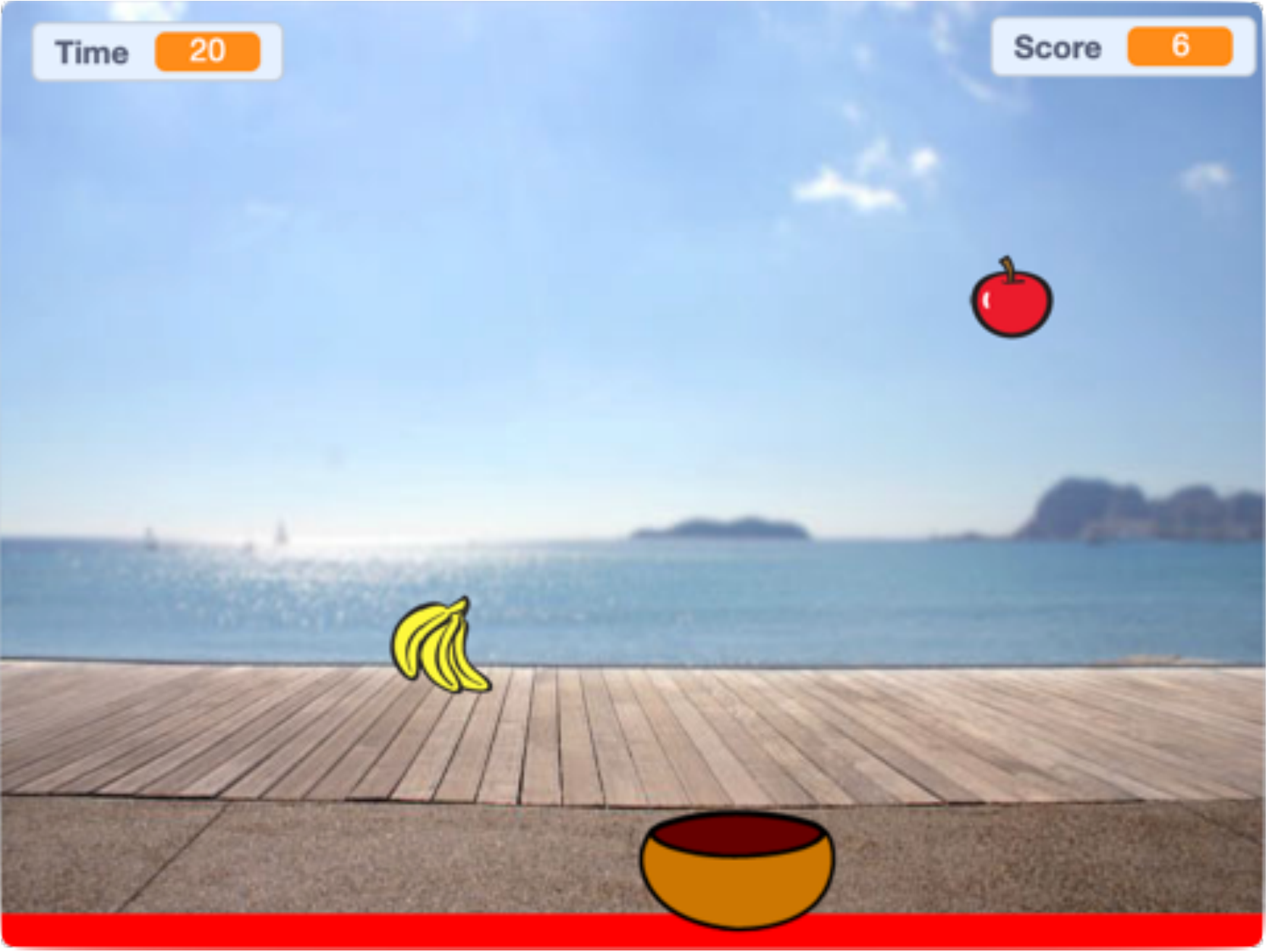}
    \caption{FruitCatching example game.}
    \label{fig:FruitCatching}
  \end{subfigure}
  \begin{subfigure}[b]{\columnwidth}
      \centering
    \includegraphics[width=\columnwidth]{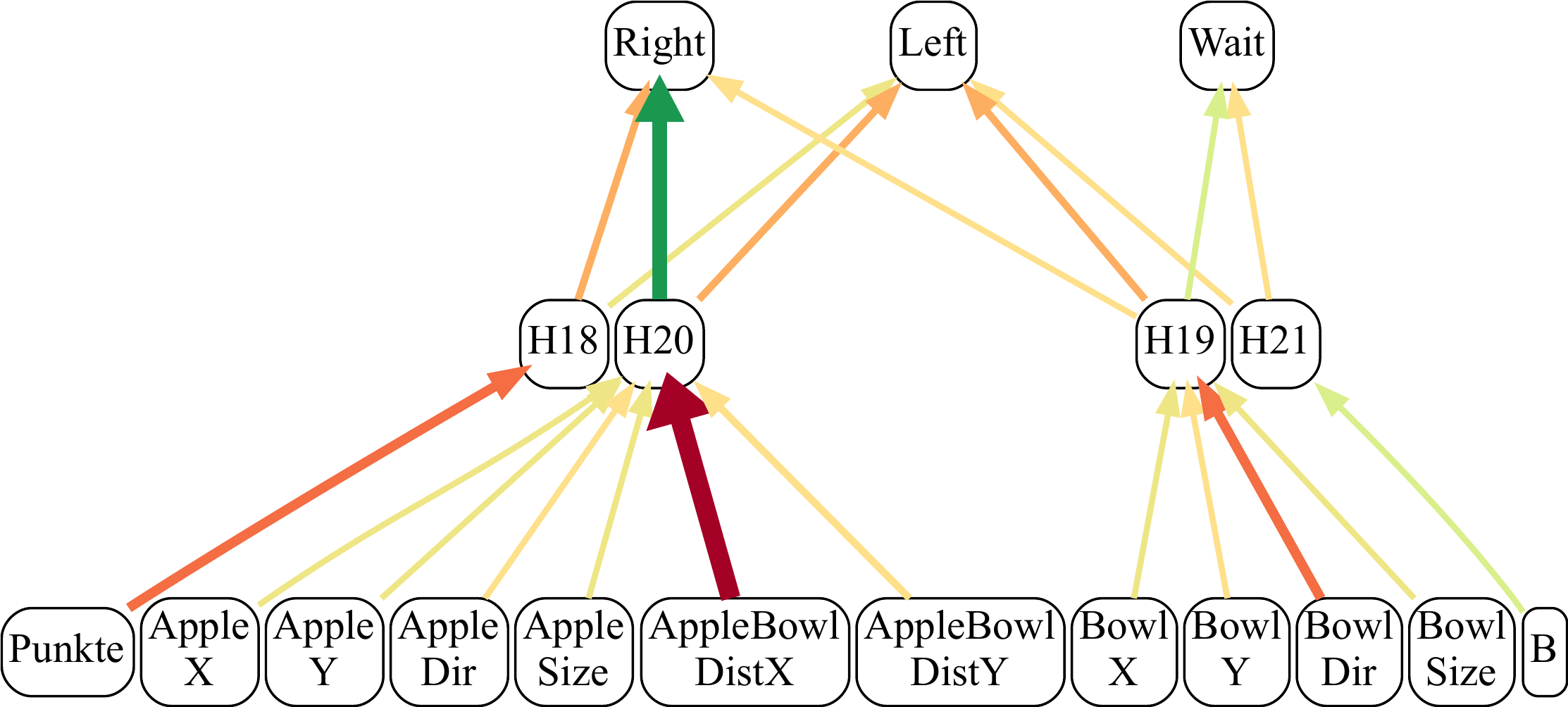}
    \caption{Example of an optimized network with excluded regression head. The width and brightness of connections represent their importance.}
    \label{fig:FruitCatching-Network}
  \end{subfigure}
  \caption{FruitCatching game and corresponding neural network that is capable of winning the game.}
\end{figure}

Game-like programs allow players to interact with them using specific input actions, and challenge them to solve dedicated tasks while enforcing a set of rules.
Once a player successfully completes the task of a game, the game usually notifies the player of his/her victory, leading to an advanced program state (\emph{winning state}).
Nowadays, game-like programs can be found in many software engineering domains. For example, the most common category of mobile apps in Google's
Playstore\footnote{[August 2022] https://www.statista.com/statistics/279286/google-play-android-app-categories/} as well as in Apple's App
Store\footnote{[August 2022] https://www.businessofapps.com/data/app-stores/} are games by
far, and games are equally common in web and desktop applications. 
Programming education is also heavily reliant on learners building
game-like programs in order to engage them with the challenges of programming.
\cref{fig:FruitCatching} shows the \emph{FruitCatching} game, a typical program created by young learners of
the \Scratch programming language~\cite{Scratch}: The player has to control a
bowl at the bottom of the screen using the cursor keys to
catch as much dropping fruit with the bowl as possible within a given time
limit.

Games are intentionally designed to be challenging for human
players in order to keep them hooked and entertained. Unfortunately, precisely
this aspect also makes game-like programs very difficult to test, and classical
approaches to automated test generation will struggle to interact with games in
a successful way \cite{deiner2022automated}. In \cref{fig:FruitCatching}, a test generator would need to
successfully catch all apples for 30 seconds without dropping a single one in order to win the game and reach the winning state.
%

While winning the game is out of question for a traditional automated test
generator, it may nevertheless succeed in producing tests for partial aspects
of a game, such as catching an individual banana or apple for the game shown in
\cref{fig:FruitCatching}. Such tests typically consist of static, timed input
sequences that are sent to the program under test to reproduce a certain
program execution path. This causes a problem: Games are inherently randomized,
and playing the same game twice in a row may lead to entirely different
scenarios. It may be possible to instrument the test execution environment such
that the underlying random number generator produces a deterministic
sequence~\cite{FlakyTests}. The \Whisker~\cite{Whisker} test framework for
\Scratch, for example, allows setting the random number generator to a fixed
seed, such that the fruit in \cref{fig:FruitCatching} always appears at the
same location in the same program. However, even the slightest change to the
program may affect the order in which the random number generator is queried, thus
potentially invalidating the test. For example, if the banana sprite is removed
from the game in \cref{fig:FruitCatching}, even though the code of the apple
sprite is unchanged the position at which the apple spawns changes, as the
order in which the random numbers are consumed in the program is different, and as a result a
test intended to catch the apple sprite would likely drop the apple. Even if by
chance the test would succeed in catching the apple despite the randomness, the
exact program state would likely be different, and an assertion on the
exact position of the apple or bowl in \cref{fig:FruitCatching} would fail,
rendering the use of classical test assertions impossible.



\begin{figure}[!tbp]
    \centering
    \includegraphics[width=\columnwidth]{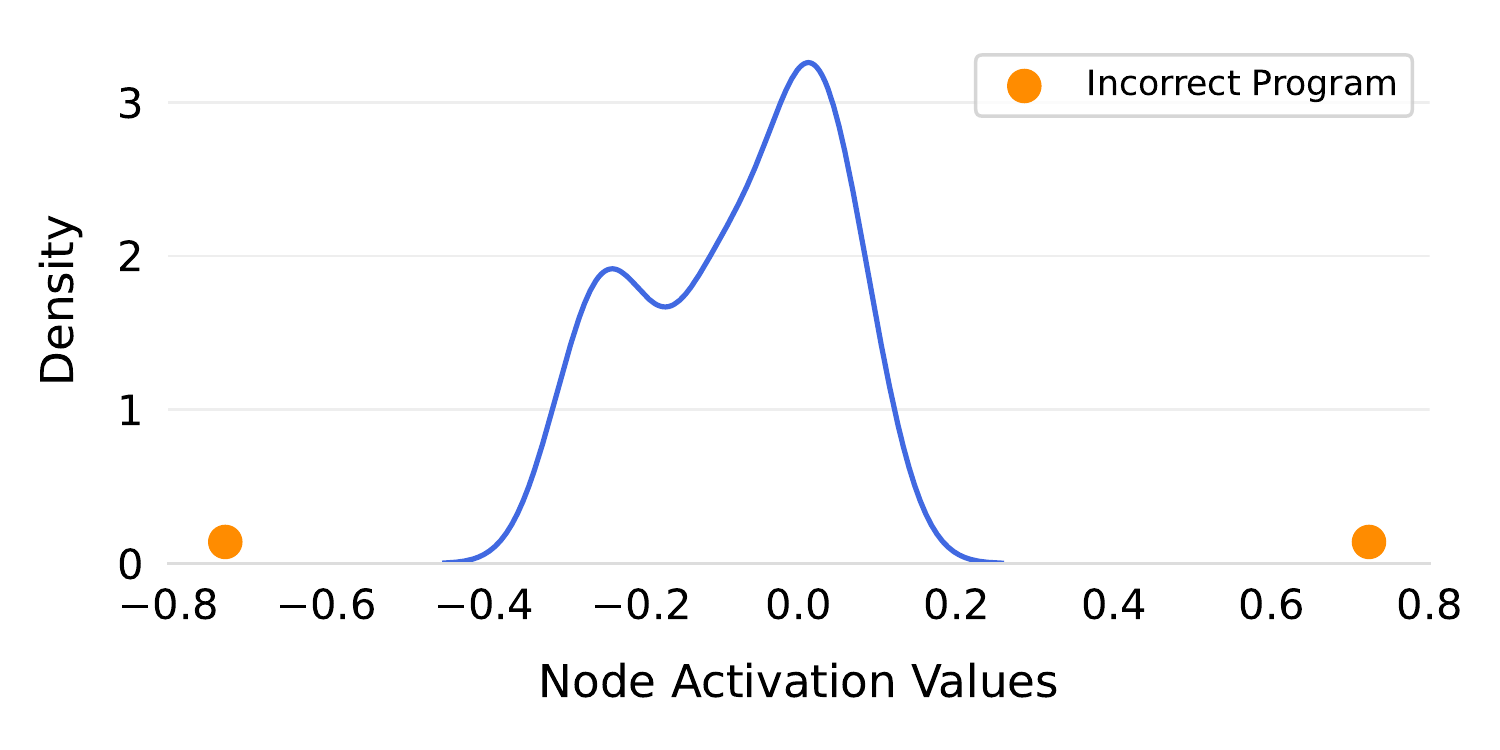}
    \caption{Activation distribution of node \emph{H20} from Fig. 1b 
at time-step 50 after 100 \emph{FruitCatching} execution. Orange dots correspond to activations from two faulty program versions.}
    \label{fig:SA}
\end{figure}

To tackle the test generation and oracle problem for game-like programs, we
propose the use of neural networks as \emph{dynamic} tests. For instance,
\cref{fig:FruitCatching-Network} shows a network structure adapted to reaching
code related to winning the \emph{FruitCatching} game
(\cref{fig:FruitCatching}). It uses aspects of the program state such as sprite
positions as input features, and suggests one of the possible ways to interact
with the program. The architecture of this example network puts strong emphasis on the
horizontal distance between the bowl and the apple which reflects that catching
apples rather than bananas is necessary to win the game, and indeed the network
is able to win the \emph{FruitCatching} game by catching all apples for 30s. In
contrast to conventional test suites consisting of static input sequences, such
dynamic tests are capable of reacting to previously unseen program states,
which is particularly useful for game-like programs that are heavily
randomized. The same dynamic tests can also serve as test oracles in a regression testing scenario by 
comparing node activations of a supposedly correct program version against a modified one.
For instance, \cref{fig:SA} shows
the activation distribution of node \emph{H20} from \cref{fig:FruitCatching-Network} at time-step 50 when 
executed 100 times on a correct version of \emph{FruitCatching}. Since the most influential input of node \emph{H20} is the horizontal distance between the bowl and the apple, the network's effort to catch the fruit can clearly be observed from the high density around the node activation of zero.
The two example incorrect program versions (mutants) depicted as orange dots disallow the player to move to the left or right, respectively, resulting in suspicious activation values that will be reported as invalid program behavior.

In this paper we introduce \ENEAT, an approach that implements this idea using 
\emph{neuroevolution}~\cite{Neuroevolution} to automatically generate
artificial neural networks that are able to test games reliably, such as the
network shown in Fig. 1b. 
\ENEAT targets games created
in \Scratch~\cite{Scratch}, a popular block-based programming environment with
over 90 million registered users and even more shared projects at the time of
this writing\footnote{[August 2022] https://scratch.mit.edu/statistics} and is
implemented on top of the \Whisker~\cite{Whisker} testing framework for
\Scratch.
Given a \Scratch program, \Neatest iteratively targets statements based on the
control dependence graph of the program, and evolves networks (i.e., dynamic
tests) to reach those statements guided by an adapted reachability-related
fitness function.
Notably, by including statements related to the winning state of a game, the
networks implicitly learn to meaningfully play and eventually win those games
without requiring any domain-specific knowledge about the games.
Using the dynamic tests it produces, \Neatest introduces a novel test oracle
approach based on \emph{Surprise Adequacy}~\cite{SA} to detect incorrect
program behavior by analyzing the networks' node activations during the
execution of a program under test.

In detail, the contributions of this paper are as follows:
\begin{itemize}
	\item We propose \ENEAT, a novel test generation approach that uses the
\emph{NeuroEvolution of Augmenting Topologies} (NEAT)~\cite{NEAT} technique to
evolve artificial neural networks capable of testing game-like programs reliably.
	\item Our testing tool implements this approach in the context of the
\Scratch educational programming environment and the \Whisker test generator
for \Scratch.
	\item We empirically demonstrate that the proposed framework is capable of producing tests for 25 Scratch games that are robust against randomized program behavior.
	\item We empirically show through mutation analysis on 25 Scratch games that the behavior of neural networks can serve as a test oracle and detect  erroneous program behavior.
\end{itemize}

\section{Background}\label{Background}

In this paper we combine the ideas of search-based software testing with
neuroevolution, considering the application domain of games implemented in the
\Scratch programming language.


\subsection{Search-based Software Testing}\label{sec:SBST}

Search-based software testing (SBST) describes the process of applying
meta-heuristic search algorithms, such as evolutionary algorithms, to the task
of generating program inputs. The search algorithm is guided by a fitness
function, which typically estimates how close a test is towards reaching an
individual coverage objective, or achieving 100\% code coverage. The most
common fitness function derives this estimate as a combination of the distance
between the execution path followed by an execution of the test and the target
statement in the control dependence graph (\emph{approach
level}~\cite{wegener2001evolutionary}), and the distance towards evaluating the
last missed control dependency to the correct outcome (\emph{branch
distance}~\cite{korel1990automated}). The approach is applicable to different
testing problems by varying the representation used by the search algorithm.
For example, SBST has been used to evolve parameters for function
calls~\cite{lakhotia2010austin}, sequences of method
calls~\cite{baresi2010testful,fraser2011evosuite,tonella2004evolutionary},
sequences of GUI interactions~\cite{mao2016sapienz,gross2012search}, or calls
to REST services~\cite{arcuri2019restful}. The search only derives test inputs
to exercise a program under test, and test oracles are typically added
separately as \emph{regression assertions}, which capture and assert the
behavior observed on the program from which the tests are
derived~\cite{xie2006augmenting,fraser2011mutation}. Since the tests derived
this way cannot adapt to different program behavior, we call them \emph{static}
tests.

\subsection{Neuroevolution}\label{sec:Neuroevolution}


Artificial neural networks (ANNs) are computational models composed of several
interconnected processing units, so-called \emph{neurons}. An ANN can be
interpreted as a directed graph where each node corresponds to a \emph{neuron
model}, which usually implements a nonlinear static \emph{transfer
function}~\cite{TransferFunctions}. The edges of the graph, which connect nodes
with each other, are represented by weights that determine the connection
strength between two neurons. Altogether, the number and type of neurons and
their interconnections form an artificial neural network's \emph{topology} or
\emph{architecture}.

Before an ANN can solve specific problems, it has to be trained in the given
problem domain. During training, \emph{learning algorithms} are used to find
suitable network parameters, such as the weights of the connections within a
network. Learning algorithms commonly apply
\emph{backpropagation}~\cite{Backpropagation}, which updates the weights by
calculating the gradient for each connection individually.
However, backpropagation on its own is only capable of tuning the weights of a neural network. 
Developing an appropriate network architecture remains a tedious task for the developer.
A promising approach to eliminate these time-consuming tasks is to apply evolutionary algorithms that can optimize both, the weights of networks and their architecture.

Instead of using a conventional learning algorithm, \emph{neuroevolution}
encodes ANNs in genomes and gathers them in a \emph{population}. In each
iteration, a subset of the current population is selected and evolved using
\emph{mutation} and \emph{mating} procedures inspired by the principles of
Darwinian evolution. Then, each genome's performance is evaluated using a
\emph{fitness function}, which reflects how close a given genome is to solving
the problem at hand. Finally, following the paradigm of survival of the
fittest, a new generation of the population is formed by selecting the fittest
genomes of the population. By exploring the search space through many
generations of mutating and mating these genomes, it is expected that
eventually a neural network with optimal topology and weights for the given
problem will be found. With this learning approach the weights, the number of
neuron layers, and the interconnections between them, can be genetically
encoded and co-evolved at the same time.

The popular \emph{NeuroEvolution of Augmenting Topologies}~\cite{NEAT} (NEAT)
algorithm solves several challenges of \emph{Topology and Weight Evolving
Artificial Neural Networks}, such as the \emph{competing conventions} problem
\cite{CompetingConventions}. NEAT solves this problem by introducing innovation
numbers that enable the comparison of genes based on their historical origin.
Another crucial aspect of NEAT is the use of
\emph{speciation}~\cite{MultimodalFunctionOptimisation} to protect innovative
networks. This is necessary since innovative topologies tend to initially
decrease the fitness of a network, leading to architectural innovations having
a hard time surviving even a single generation. Speciation ensures that neural
networks with similar structures have to primarily compete against other
networks in the same species, which gives them time to optimize their weights.
Two networks are assigned to the same species if the number of mismatching
genes and the average connection weight difference falls below a threshold,
which is adjusted in each generation of the algorithm based on the number of
species that are present in the current generation: if there are fewer species
than desired, the threshold is reduced (e.g., by 0.3) and vice versa.

NEAT's mutation operators can be divided into two groups: \emph{structural
mutations} and \emph{non-structural mutations}. The former modify a 
network's topological structure, while the latter operate on the attributes of
connection genes. The crossover operator randomly selects two parents and
aligns their connection genes using the assigned innovation numbers. Then, both
connection gene lists are sequentially traversed, and certain rules are applied
to each gene pair: Similar genes are either inherited randomly from one parent
or combined by averaging the connection gene weights of both parents. On the
other hand, genes with differing innovation numbers are always inherited from
the more fit parent.

\subsection{Testing Scratch Programs}\label{sec:ScratchTesting}


The block-based programming language \Scratch~\cite{Scratch} provides
predefined command blocks encoding regular programming language statements as
well as domain-specific programming instructions that make it easy to create
games. \Scratch programs are created by visually arranging these command blocks
according to the intended program behavior. 
%
\Scratch programs consist of a \emph{stage} and a collection of \emph{sprites}.
The former represents the application window and displays a background image
chosen by the user. Sprites, on the other hand, are rendered as
images on top of the stage and can be manipulated by the user. Both the stage
and the sprites contain \emph{scripts} formed by combining command blocks to
define the functional behavior of the program. Besides control-flow structures
or variables, the blocks represent high-level commands to manipulate sprites.
\Scratch programs are controlled via user inputs such as key presses, text, or
mouse movement. To engage learners, the \Scratch programs they create are
usually interactive games.

Testing and dynamic analyses are important for \Scratch programs:
Even though the \Scratch blocks ensure syntactic correctness,
functional errors are still feasible, raising the need for tests to validate
the desired behavior of programs. In particular, tests are a prerequisite for
effective
feedback~\cite{keuning2016towards,buffardi2015reconsidering,yi2017feasibility,gulwani2018automated} and hints~\cite{kim2016apex} to learners, as well as for
automated assessment~\cite{ala2005survey,ihantola2010review}. 
\Whisker~\cite{Whisker} is a framework that
automates testing of \Scratch programs by sending inputs to \Scratch programs
and comparing the system's observed behavior against a predefined specification.

In order to discover as many system states as possible and thereby enhance the
chances of finding violated properties, it is crucial to have a widespread set
of test inputs. The \Scratch inputs used to stimulate the given program under test
are assembled in the so-called \emph{test harness}. \Whisker offers two types
of test harnesses: a manually written test harness
and an automatically generated test harness using search-based test
generation approaches \cite{deiner2022automated}.

Conventional search-based test generation is often not powerful enough to cover advanced
program states, especially in game-like projects~\cite{deiner2022automated}. In order to increase the
effectiveness of automatic test generation, we introduce a novel test generation
approach by extending the \Whisker testing tool. The proposed test generator uses
neuroevolution to produce neural networks that are capable of extensively exercising
game-like programs.

\section{Testing with Neuroevolution}\label{sec:TestingScratch}

Traditional tests of static input sequences cannot react to deviating program behavior properly and therefore fail to test randomized programs reliably.
As a robust alternative to static test suites, we introduce \Neatest, an extension of the \Whisker test generation framework \cite{Whisker} that is based on the NEAT \cite{NEAT} algorithm.
\Neatest generates dynamic test suites consisting of neural networks that are able to withstand and check randomized program behavior.

Algorithm~\ref{alg:Neatest} shows the test generation approach of \Neatest,
which starts in line 5 with the selection of a target statement $s_t$ based on
the CDG and a list of yet uncovered program statements
(\cref{sec:SelectTargetStatement}). Next, a new population of
networks is created depending on the search progress: If no statements have been
covered yet, the first population consists of networks where
each sprite's feature group (\cref{sec:FeatureExtraction}) is fully connected
to a single hidden node that is connected to all output nodes. Otherwise,
a new population is produced by cloning and mutating networks that proved to be
viable solutions for previously targeted statements. By selecting prior
solutions, we focus on networks already capable of reaching interesting program states,
which may be prerequisites for reaching more advanced states.
To avoid unnecessary complex network structures, a certain share (e.g., 10\%) of
every new population consists of freshly generated networks. 
We require hidden layer neurons to determine the correctness of a program (\cref{sec:ActivationTraces}).

The following loop over all networks in line 9 starts with the \emph{play loop} in which a network simulates a human player by repeatedly sending input events to a given \Scratch program (\cref{sec:Playing}).
The inputs for the networks are extracted from the current state of the program, while the possible outputs are determined by the actions a player can take in the game (\cref{sec:FeatureExtraction}).

After a network has finished its playthrough, lines 11 to 16 evaluate
how reliably the network can cover the current target statement $s_t$
(\cref{sec:NetworkFitness}). If no network manages to cover $s_t$ with sufficient robustness (line 17), the algorithm executes line 24 and evolves a new generation of
networks using the NEAT algorithm~(\cref{sec:Neuroevolution}). This new
generation is then the starting point for the next iteration of the algorithm.
Once a network has been optimized to cover $s_t$ reliably for a user-defined
number of times $r_d$, the network is added as a test to the dynamic test
suite, and the next target statement is selected (lines 18--20). This process is
repeated until the search budget is exhausted or all statements have been
covered, and the dynamic test suite is returned to the user. The dynamic tests
serve not only to exercise the programs, but also as test oracles for
validating the functionality in a regression testing scenario
(\cref{sec:ActivationTraces}).

\begin{algorithm}[tb]
\SetKwInOut{Input}{input}
\SetKwInOut{Output}{output}
\SetKwProg{Fn}{function}{:}{end}
\SetFuncSty{textsf}
  \Input{control dependence graph ~$CDG$}
  \Input{list of uncovered program statements ~$S$}
  \Input{desired robustness count ~$r_d$}
  \Output{dynamic test suite~$D$}
	\Fn{\Neatest({$CDG$, $S$, $r_d$})}{
		$ requireNextStatement \gets true$\;
		\While{stopping condition not reached} {
			\If{$requireNextStatement$}{
			$ s_t \gets$ selectTargetStatement($CDG$, $S$)\;
			$ P \gets$ generatePopulation($D$)\;
			$ requireNextStatement \gets false$\;
			}
			\ForEach{network \( n \in P \)}{
				initiatePlayLoop($n$)\;
        		$f \gets $ calculateFitness($n$)\;
        		\If(\tcp*[f]{Statement was covered}){$f == 1$}{
        			$r_c \gets $ robustnessCheck($n, r_d$)\;
        			$f \gets f + r_c$
        			
        		}
        		$network.fitness \gets f$\;
        		\If{$f == r_d$}{
        			$D \gets D + n$\;
        			$S_u \gets S_u - s_t$\;
        			$requireNextStatement \gets true$
        		}
      		}
      		\If{$!requireNextStatement$}{
        		$P \gets$ NEAT($P$)\;
        	}
		}
  		\Return{$ D $}
  		}
    \caption{\ENEAT}
    \label{alg:Neatest}
\end{algorithm}

\subsection{Explorative Target Statement Selection}\label{sec:SelectTargetStatement}

\begin{figure}[!tbp]
  \begin{subfigure}[b]{\columnwidth}
    \centering
    \includegraphics[width=.8\columnwidth]{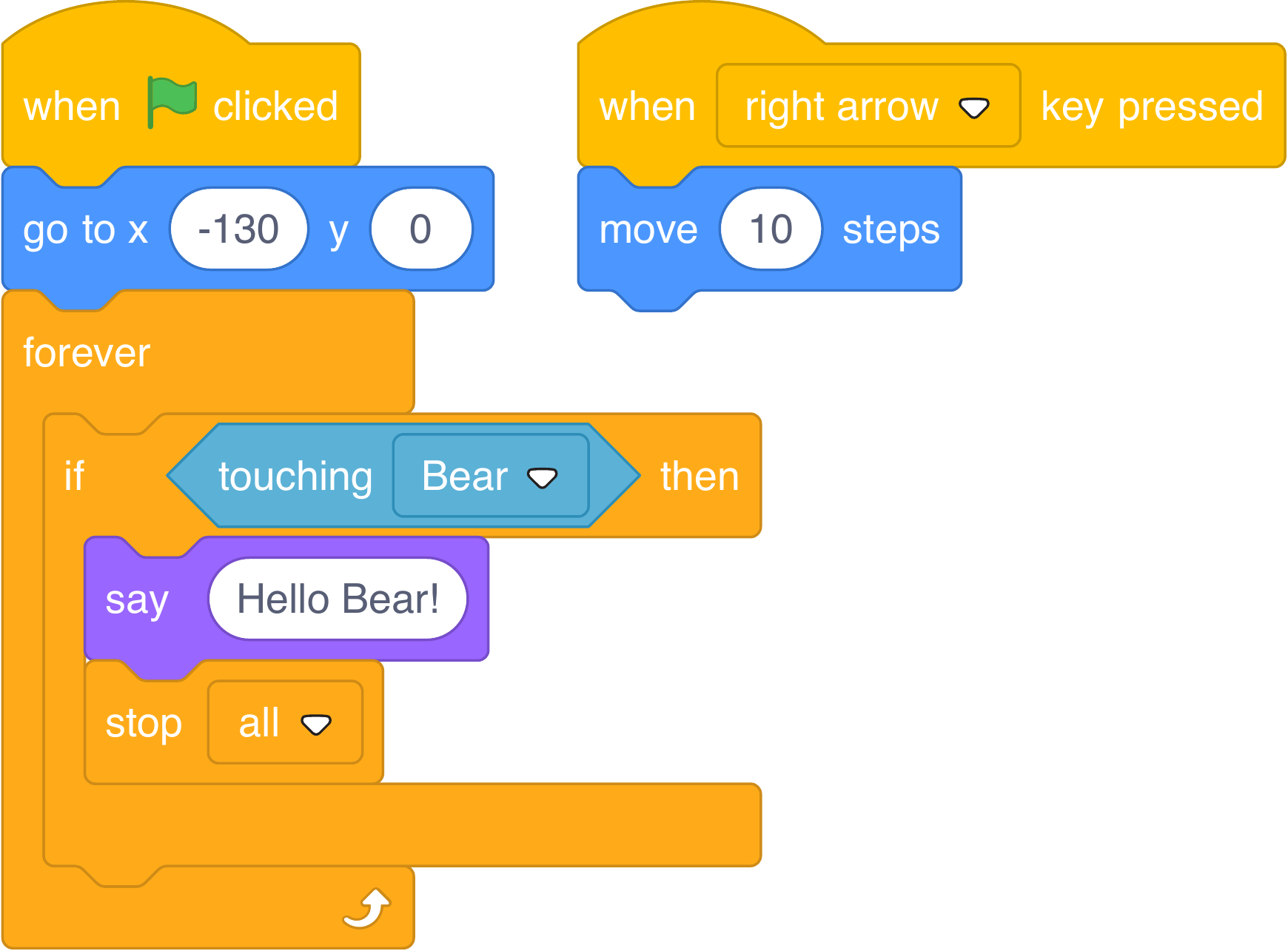}
    \caption{Two \Scratch scripts that implement the movement of the hosting sprite and a check if the hosting sprite touches a bear sprite.}
    \label{fig:CDG-Scratch}
  \end{subfigure}
  \begin{subfigure}[b]{\columnwidth}
      \centering
    \includegraphics[width=.8\columnwidth]{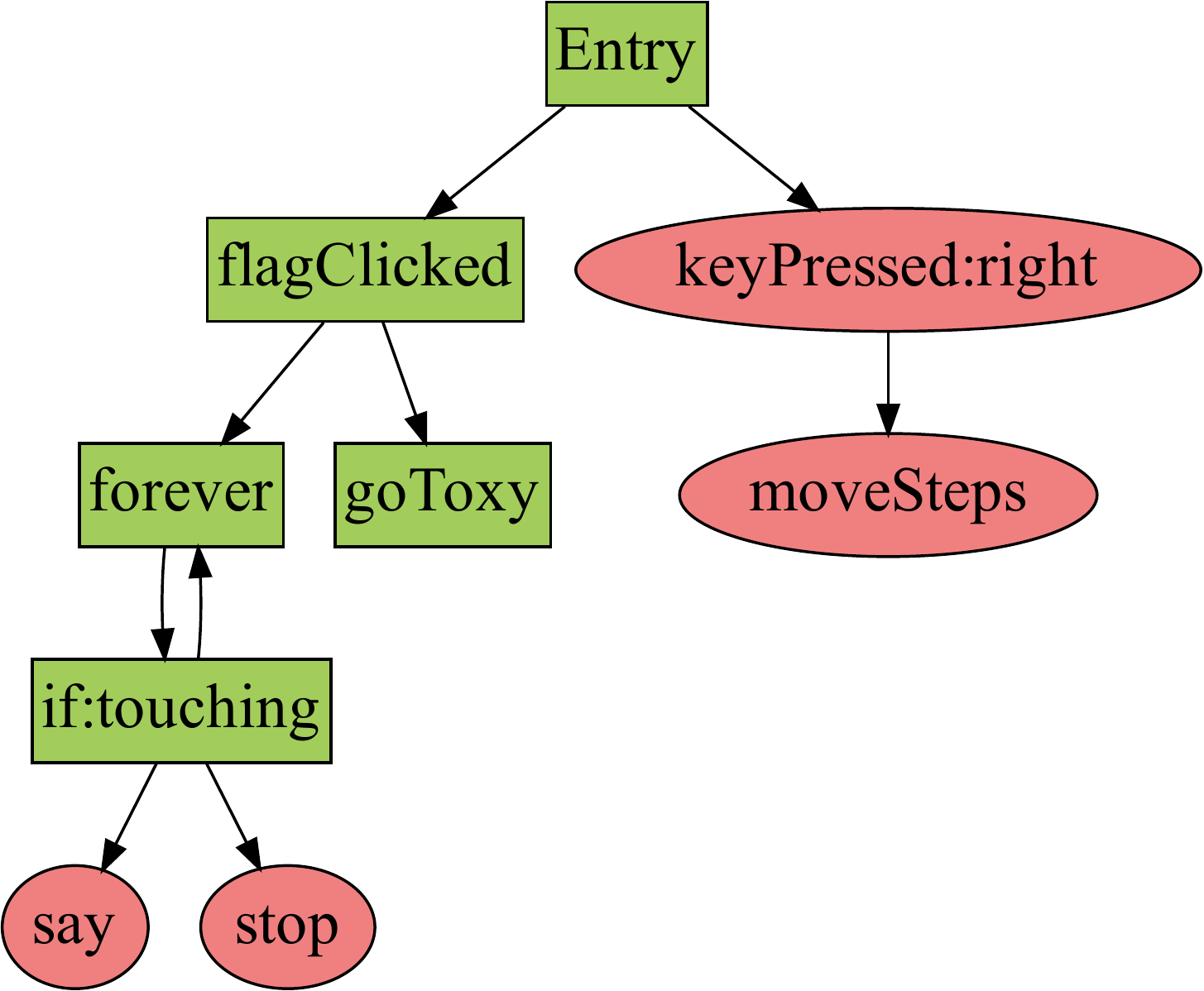}
    \caption{CDG for the depicted script with squared green and oval red nodes representing covered and uncovered \Scratch statements.}
    \label{fig:CDG-Graph}
  \end{subfigure}
  \caption{Example of a \Scratch script and the corresponding CDG.}
  \vspace{-1.5em}
\end{figure}

\Neatest iteratively selects increasingly more difficult target statements 
to guide the creation of incrementally more complex networks.  
In order to achieve this, targets are selected from the test program's control
dependence graph (CDG) \cite{CDG}, a directed graph that shows which
program statements depend on which control locations. Since \Scratch programs
typically contain many small scripts that communicate via events and messages,
we combine the scripts of a \Scratch program into an interprocedural CDG that covers the control flow of the entire
program.
For example, \cref{fig:CDG-Scratch} shows a \Scratch program consisting of two scripts, which are combined into one interprocedural CDG~(\cref{fig:CDG-Graph}) by introducing an artificial \emph{Entry} node.

 Target statements $s_t$ are selected from children of already covered
program statements, as advanced states can only be reached if previous control
locations can be passed appropriately. If no statements have been covered yet, the root node of the CDG (\emph{Entry}) is selected.
In case more than one viable child exists, \Neatest favors program statements
that have accidentally been reached while focusing on other program states but
have not yet passed the robustness check. 
Considering the example of \cref{fig:CDG-Graph} and assuming that the search was able to trigger the right key-press once without covering the corresponding event handler reliably, the next target statement would be set to the \emph{keyPressed:right} statement.
Otherwise, if none of the three statements that have a directly covered parent (\emph{say}, \emph{stop}, \emph{keyPressed:right}) had been reached once, $s_t$ would be randomly set to one of these three statements.
To avoid getting stuck trying to cover hard-to-reach statements while other,
perhaps easier-to-reach states are still present, \Neatest changes its
currently selected target after a user-defined number of consecutive generations without improvement of fitness values. 

\subsection{Play Loop}\label{sec:PlayingLoop}

Within the Play Loop, the networks govern the execution of a program by repeatedly sending input events until the execution halts, a pre-defined timeout has been reached, or the targeted statement $s_t$ was covered. 
To make a decision on which user event to select at a given moment, we provide the networks with input features extracted from the current program state.
The selected event is then sent to the \Scratch program to advance the program execution.

\subsubsection{Feature Extraction}\label{sec:FeatureExtraction}

Input features are fed to the input neurons of networks and play a crucial role during a playthrough by serving as the neural networks' eyes. 
The features are collected by filtering attributes of visible sprites by their relevance to the given \Scratch program, as determined by analyzing the source code:
\begin{itemize}
\item \textbf{Position} of a sprite on the \Scratch canvas is always added and defined through its x- and y-coordinates in the range $x \in [-240, 240]$ and $y \in [-180, 180]$.
\item \textbf{Size} of a sprite is always added. The size bounds depend on the currently selected \emph{costume} (i.e., visual representation).
\item \textbf{Heading direction} iff the rotation style is set to \emph{all around}. Defined through an angle $\alpha$ in the range $\alpha \in [-180, 180]$.
\item \textbf{Selected costume} iff the sprite contains code to change its costume. Defined as $i \in \mathbb{N}$, indexing the current costume.
\item \textbf{Distance to sensed sprite} iff the sprite contains code that checks for touching another sprite. Defined through the x and y distance to the sensed sprite in the range $[-600, 600]$.
\item \textbf{Distance to sensed color} iff the sprite contains code that checks for touching a given color. Colors are measured by four rangefinders, which measure the distance on the canvas to the sensed color in the directions $[0, 90, 180, -90]$, with 0 representing the current \emph{heading direction} of the sprite. Admissible values are again restricted to $[-600, 600]$.
\end{itemize}

In addition to relevant sprite attributes, we also add all program variables, for example, the current score of a game, to the set of input features.
We add input nodes for all extracted features, and group them by the sprite they belong to.
Since we only include relevant features for which corresponding code exists, the input features may not only 
differ between games but also within the same game, for
example whenever a sprite creates a clone that has its own attributes. However, neuroevolution allows
the framework to adapt to these input feature changes dynamically by
adding input nodes to the networks whenever a new behavioral pattern is
encountered. Finally, to ensure all extracted input features have the same
weight, they are normalized to the range $[-1, 1]$.

Output features of a network resemble the set of user events a human player can
send to the \Scratch application at a given point in time. The framework
simulates human inputs using the following \Scratch events:
\begin{itemize}
\item \textbf{KeyPress}: Presses a specific key on the keyboard for a parameterized amount of time.
\item \textbf{ClickSprite}: Clicks on a specific sprite.
\item \textbf{ClickStage}: Clicks on the stage.
\item \textbf{TypeText}: Types text using the keyboard.
\item \textbf{MouseMove}: Moves the mouse to a parameterized position.
\item \textbf{MouseMoveTo}: Moves the mouse to a position inferred by static analysis (e.g., location of another sprite).
\item \textbf{MouseDown}: Presses the left mouse button for a parameterized amount of time.
\item \textbf{Sound}: Sends simulated sound to the program.
\item \textbf{Wait}: Waits for a parameterized amount of time.
\end{itemize}
\Neatest analyzes the source code of all actively running scripts for a given
program state in order to select the subset of event types for which active
event handlers exist. 
After minimizing the set of event types, we create one output node for each remaining event.
In case there are
parameterized events, such as \emph{MouseMove} events requiring coordinates where the
mouse pointer should be moved to, the classification problem is further
extended to a combination of a classification and regression problem. These  parameters are produced via a multi-headed
network model~\cite{Multi-Headed-Network} in which regression
nodes are mapped to parameterized events. Similar to input features, the set of
feasible output features and the number of required output nodes may change
throughout the course of a game.

\subsubsection{Network Activation \& Event Selection}\label{sec:Playing}

The problem of deciding which action to take at the current state of a game is
cast to a multi-class single-label classification problem. 
Classification steps start by loading the input features of the current program state into the corresponding input nodes. 
Then, the neural network is activated in order to produce an output based on the fed input features.
During network activation, values residing in the input nodes
flow through the network in several timesteps. In every timestep, each node
adds up all the activation values from the previous timestep and calculates the result of
its own activation using the \emph{tanh} function \cite{Tanh} on the summed-up values.

To not perturb values gathered during \emph{feature extraction}, all input
nodes send the normalized input features directly into the network
without activation functions. In a single timestep, activation
only flows from one neuron to the next, thus several network activations are required for
the input nodes' signals to reach the output nodes. 

After the input signal has reached the output nodes, a probability
distribution is calculated over the set of available events using the traditional \emph{softmax
function}~\cite{Softmax}.
The classification task is extended into
a regression problem whenever a network selects a parameterized event. Parameters for the selected event are gathered by querying the responsible
nodes in the network's regression head. 

Selected events are passed on to the \Scratch VM, where they may trigger the activation of previously inactive scripts. 
Upon receiving an input event, the \Scratch VM executes a \Scratch step by sequentially executing blocks of currently active scripts until either a block that halts the program execution or the end of the script are reached. 
The \Scratch step finishes by updating the state of the VM to match the effects of the executed blocks.
The playing loop then starts anew by extracting the input and output features of the updated VM.

\subsection{Network Evaluation}\label{sec:NetworkFitness}

Given a selected target statement $s_t$, the evolution of networks is guided by a fitness function $f$ which estimates how close a network is to reaching $s_t$ and, once reached, how reliably it is executed.
We first calculate the objective function $f_{s_t}$~\cite{deiner2022automated} that guides the search towards reaching $s_t$ using a linear combination of:
\begin{itemize}
	\item \textbf{Approach level (AL)}: Integer defining the number of control dependencies between the immediate control dependency of the target and the closest covered statement.
	\item \textbf{Branch distance (BD)}: Indicates how close a network is to satisfying the first missed control dependency. Zero if the control location has been passed towards the target statement.
		\Scratch often allows for a precise definition of the branch distance. For example, in \cref{fig:CDG-Graph}, the branch distance for the \emph{if:touching} statement is defined as the normalized distance between the corresponding sprites. 
	\item \textbf{Control flow distance (CFD)}: As execution may stop in the middle of a sequence of blocks (e.g., because of timed statements), the CFG counts the number of blocks between the targeted statement and the nearest covered statement.
\end{itemize}
To avoid deceiving fitness landscapes, we normalize the branch distance and the control flow distance in the range [0, 1] using the normalization function $\alpha(x) = x/(1+x)$~\cite{arcuri_it_2013}, and multiply the approach level with a factor of two:
\setlength{\belowdisplayskip}{0.2em}
\setlength{\abovedisplayskip}{0.2em}
\begin{equation*}
  f_{s_t}  = 2 \times AL + \alpha(BD) + \alpha(CFD)
\end{equation*}
In case $f_{s_t} > 0$, the target statement was not covered and we set the
fitness to $f = 1/f{s_t}$ to create a maximization objective towards covering $s_t$. Otherwise ($f_{s_t} = 0$), the network managed to cover the target
once and we evaluate whether it can also cover the same target in 
randomized program scenarios. For this purpose, we generate up to $r_d - 1$ random seeds, count for how many seeded executions $r_c$
the network is able to reach the target statement again, and finally set the fitness function to $f = 1 + r_c$. We decrement the desired robustness count $r_d$ and increment the fitness value $f$ by one to account for the first time $s_t$ was covered  due to $f_{s_t} = 0$.


The fitness function ensures that networks that are closer or more robust in
covering a specific statement are prioritized during the evolution. If a
network eventually passes the robustness check ($f = r_d$), the network is
added to the dynamic test suite as test for target statement $s_t$ and the next
target is selected by querying the CDG as described in
\cref{sec:SelectTargetStatement}. Since covering one statement usually leads to
reaching other related statements as well, we simultaneously evaluate whether
the network may also satisfy the robustness check for other yet
uncovered program statements.

\subsection{Neural Networks as Test Oracles}\label{sec:ActivationTraces}

In order to determine whether a test subject is erroneous, we observe how surprised a network behaves when confronted with the program under test and regard highly surprising observations as suspicious.
As a metric, we use the \emph{Likelihood-based Surprise Adequacy} (LSA) \cite{SA}, which
uses kernel density estimation \cite{KDE} to determine how surprising the activations of selected neurons are when the network is presented with novel inputs.
In the application scenario of regression testing, node activations originating from a supposedly correct program version serve as the ground truth and will be compared to activations occurring during the execution of the test subject. 
To incorporate many randomized program scenarios, we collect the ground truth activations by executing the networks on the sample program using a series of different random seeds.
Defective programs are then detected if the LSA value surpasses a pre-defined threshold, or for the special case that all ground truth traces have a constant value, if the surprise value is above zero since in this scenario already tiny deviations are suspicious.

\Neatest only evaluates activation values occurring within hidden nodes because these nodes strike a nice balance between being close to the raw inputs that represent the program state while already offering an abstraction layer by incorporating a network's behavior.
This allows differentiating between suspicious and novel program behavior, an essential characteristic for testing randomized programs that is missing in traditional static test assertions.
For example, in the \emph{FruitCatching} game (\cref{fig:FruitCatching}), the apple and bananas may spawn at entirely new x-coordinates, but the program state is still correct as long as they spawn at a certain height.
Successfully trained networks can differentiate between novel and suspicious states because they are optimized to catch falling fruit by minimizing the distance between the bowl and the fruit, as shown by the zero-centered activation distribution in \cref{fig:SA}.
Since both example defective programs, represented as orange dots, are far off the distribution center, they are rather surprising, with LSA values of 50 and 102.
Therefore, given a threshold below these values, \Neatest correctly classifies these program versions as faulty.

In contrast to the original definition of LSA~\cite{SA}, we do not compare
groups of nodes that may reside in the same layer but compare the activation
values of hidden nodes directly. Furthermore, we classify activation values by
the \Scratch step in which they occurred. This precise approach of comparing
node activations offers two major advantages: First, we can detect tiny but
nonetheless suspicious program deviations that may only occur within a single
node by calculating the LSA value for each node separately. Second, if a
suspicious activation within a node was detected, we can identify the exact
point in time during the execution and the rough reason for the suspicious
observation by following the suspicious node's incoming connections back to the
input nodes. 

Changes to the structure of a network, which may occur during an
execution, can also provide strong evidence of suspicious program
behavior. For instance, output nodes may be added if a game can
process different user events, and input nodes may be added for new
sprite clones (cf. \cref{sec:PlayingLoop}). Therefore, we can deduce
from a change in a network's structure that the test subject exposes
diverging and maybe even erroneous program behavior.

\section{Evaluation}
In order to evaluate the effectiveness of the proposed testing framework we aim to answer the following three research questions:

\begin{itemize}
    \item \textbf{RQ1}: Can \Neatest optimize networks to exercise \Scratch games reliably?
    \item \textbf{RQ2}: Are dynamic test suites robust against randomized program behavior?
    \item \textbf{RQ3}: Can neural networks serve as test oracles?
\end{itemize}

\noindent \Neatest is integrated and available as part of the \Whisker testing framework\footnote{[August 2022] https://github.com/se2p/whisker}; the experiment dataset, parameter configurations, raw results of the experiments, and scripts for reproducing the experiments are publicly available on GitHub\footnote{[August 2022] https://github.com/FeldiPat/ASE22-Neatest-Artifact}.

\subsection{Dataset}

\begin{table}[tb]
    \centering
    \caption{Evaluation games.}
    \label{tab:Dataset}
        \vspace{-1em}
    \resizebox{\columnwidth}{!}{
        \begin{tabular}{lrrrlrrr}
            \toprule
            Project & \rotatebox{90}{\texttt{\#} Sprites} & \rotatebox{90}{\texttt{\#} Scripts} & \rotatebox{90}{\texttt{\#} Statements} &
            Project & \rotatebox{90}{\texttt{\#} Sprites} & \rotatebox{90}{\texttt{\#} Scripts} & \rotatebox{90}{\texttt{\#} Statements} \\
            \midrule
            BirdShooter        & 4  & 10 & 69  & FlappyParrot  & 2   & 7    & 37    \\
            BrainGame          & 3  & 18 & 76  & Frogger       & 8   & 22   & 105   \\
            CatchTheDots       & 4  & 10 & 82  & FruitCatching & 3   & 4    & 55    \\
            CatchTheGifts      & 3  & 7  & 68  & HackAttack    & 6   & 19   & 93    \\
            CatchingApples     & 2  & 3  & 25  & LineUp        & 2   & 4    & 50    \\
            CityDefender       & 10 & 12 & 97  & OceanCleanup  & 11  & 22   & 156   \\
            DessertRally       & 10 & 27 & 212 & Pong          & 2   & 2    & 15    \\
            DieZauberlehrlinge & 4  & 14 & 87  & RioShootout   & 8   & 26   & 125   \\
            Dodgeball          & 4  & 10 & 78  & Snake         & 3   & 14   & 60    \\
            Dragons            & 6  & 33 & 381 & SnowballFight & 3   & 6    & 39    \\
            EndlessRunner      & 8  & 29 & 163 & SpaceOdyssey & 4   & 13   & 116   \\
            FallingStars      & 4  & 4  & 91  & WhackAMole    & 10  & 49   & 391   \\
            FinalFight         & 13 & 48 & 286 & Mean          & 5.5 & 16.6 & 118.3 \\
            \bottomrule
        \end{tabular}
    }
            \vspace{-1em}

\end{table}

To establish a diverse dataset for the evaluation, we collected 187 programs that represent the educational application domain of \Scratch from an introductory \Scratch book\footnote{[August 2022] https://nostarch.com/catalog/scratch}, prior work~\cite{Whisker}, and four tutorial websites for children (Code Club\footnote{[August 2022] https://projects.raspberrypi.org/en/codeclub}, Linz Coder Dojo\footnote{[August 2022] https://coderdojo-linz.github.io/uebungsanleitungen/programmieren/scratch/}, learnlearn\footnote{[August 2022] https://learnlearn.uk/scratch/}, junilearning\footnote{[August 2022] https://junilearning.com/blog/coding-projects/}).
Three selection criteria were used to determine if a project included in this set of 187 programs should be considered for evaluation:
First, a given project has to resemble a game that processes user inputs supported by the \Whisker framework
and challenges the player to achieve a specific goal while enforcing a set of rules.
Second, the game must include some form of randomized behavior.
Finally, we exclude games that do not have a challenging winning state and can thus be easily covered just by sending arbitrary inputs to the program.
Applying all three selection criteria to the mentioned sources and skipping programs that are similar to already collected ones in terms of player interaction and the overall goal of a game resulted in a total of 25 programs (including the \emph{FruitCatching} game shown in \cref{fig:FruitCatching}); statistics of these games are shown in \cref{tab:Dataset}.
Overall, each of the 25 games offers unique challenges and ways of interacting
with the program.

\subsection{Methodology}
All experiments were conducted on a dedicated computing cluster containing nine nodes, with each cluster node representing one AMD EPYC 7443P CPU running on 2.85 GHz.
The \Whisker testing framework\footnote{[August 2022] https://github.com/se2p/whisker} allows users to accelerate test executions using a customizable acceleration factor.
In order to speed up the experiments, all conducted experiments used an acceleration factor of ten.

\vspace{0.2cm}
\noindent\textbf{RQ1:} We evaluate whether \Neatest can train networks to cover program states \emph{reliably} by comparing the proposed approach against a random test generation baseline.
While \Neatest produces dynamic tests in the form of neural networks, the random testing approach randomly selects input events from a subset of processable events and saves the chosen events as static test cases.
\Neatest maintains a population of 300 networks, and both methods generate tests until a search budget of 23 hours has been exhausted.

The combination of 300 networks per population and the search budget of 23 hours allows the framework to conduct a reasonable amount of evolutionary operations in order to sufficiently explore the search space of the most challenging test subjects in the dataset.
However, for some test subjects, satisfying results may already be found using a population size of 100 and a search budget of only one hour.
In general, the easier a winning state can be reached and the fewer control dependencies a \Scratch program has, the fewer resources are required.

We set the number of non-improving generations after which a target is switched to five because this value provides the evolution enough time to explore the search space while also avoiding wasting too much time on difficult program states.
To distribute the 300 networks of a population across the species such that each species can meaningfully evolve the weights of its networks, we set the targeted species number to ten.
For both test generators, the maximum time for a single playthrough was set to ten seconds, which translates to a play duration of up to 100 seconds due to the acceleration factor of ten.
The remaining neuroevolution hyperparameters are set according to the original NEAT study~\cite{NEAT}.

As a metric of comparison, we consider coverage values across 30 experiment repetitions with the
\emph{Vargha \& Delaney} (\atwelve) effect size~\cite{VD} and the
\emph{Mann-Whitney-U-test}~\cite{Mann-Whitney-U} with $\alpha = 0.05$ to
determine statistical significance. To evaluate whether \Neatest can train networks to reach challenging program states, we manually identified statements representing a winning state for each program and report how often this winning state was reached within the 30 experiment repetitions. Since we are
interested in tests that are robust against randomized program behavior, we
treat a statement only as covered if a given test passes the robustness check (\cref{sec:NetworkFitness}) for ten randomly seeded program
executions.

\vspace{0.2cm}
\noindent \textbf{RQ2.} In order to explore if the generated dynamic test suites are truly robust against diverging program behavior, we randomly extract for each program ten dynamic and static test suites that were produced during RQ1.
For the purpose of eliminating differences originating from the test generation process, both test suite types are extracted from the \Neatest approach.
The static test suites correspond to the input sequences successful networks produced when they entered the robustness check.
All extracted test suites are then executed on the corresponding programs using seeds different from those during the test generation phase.

We evaluate the robustness of both test suite types by reporting the difference in
coverage between the test generation and test execution phase and investigate
the effectiveness of dynamic test suites against static test suites by comparing the
achieved coverage and \atwelve values. Similar to RQ1, we use the
\emph{Mann-Whitney-U-test} to report statistically significant results. To
match the application scenario of test suites, we do not use robustness checks
and treat a statement as covered if it has been reached at least once. Within
RQ2, we compensate for random influences by applying every extracted test suite
to ten new random seeds, which results in 100 test runs for each test suite type on
every dataset project.

\vspace{0.2cm}
\noindent \textbf{RQ3.} To answer the question if networks can serve as test oracles, we extended \Whisker with a mutation analysis feature implementing the eight different mutation operators shown in \cref{tab:mutationOperators}, and applied mutation analysis to all 25 projects.
These mutation operators were selected based on the traditional set of sufficient mutation operators \cite{offutt1996experimental}.
After generating mutant programs, we execute the dynamic test suites of RQ2 on the respective mutants and measure how surprised the networks are by the modified programs.
The ground truth activation traces required for calculating the LSA are generated by executing each test suite 100 times on the unmodified program using randomly generated seeds.
A program is marked as mutant if the LSA value of a single node activation surpasses an experimentally defined threshold of 30.
Besides calculating the LSA value, mutants are also killed if the network structure changes while being executed on the mutant.

We answer RQ3 by reporting the distribution of killed mutants across the eight applied operators.
Furthermore, we verify that the networks do not simply mark every program as a mutant by calculating the false-positive rate on unmodified programs.
Similar to RQ2, we account for randomness within the mutant generation and network execution process by repeating the experiment for all extracted test suites ten times using varying seeds.
To avoid an explosion of program variations, each experiment repetition was restricted to 50 randomly selected mutants per operator.

\begin{table}[tb]
    \caption{Mutation operators.}
    \label{tab:mutationOperators}
    \vspace{-1em}
    \resizebox{\columnwidth}{!}{
        \begin{tabular}{ll}
        	\toprule
        	Operator 	& Description \\
        	\midrule
        	Key Replacement Mutation (KRM) 	& Replaces a block's \emph{key} listener. \\
        	Single Block Deletion (SBD) & Removes a single block. \\
        	Script Deletion Mutation (SDM) & Deletes all blocks of a given script. \\
        	Arithmetic Operator Replacement (AOR) & Replaces an arithmetic operator. \\
        	Logical Operator Replacement (LOR) & Replaces a logical operator. \\
        	Relational Operator Replacement (ROR) & Replaces a relational operator. \\
        	Negate Conditional Mutation (NCM) & Negates boolean blocks. \\
        	Variable Replacement Mutation (VRM) & Replaces a variable. \\
        	\bottomrule
        \end{tabular}
    }
\end{table}

\subsection{Threats to Validity}
\textbf{External Validity:} The dataset of 25 projects covers a broad range of game types and program complexities.
Even though high priority was given to establish a dataset of games with many different objectives and ways to interact with them, we cannot guarantee that the results generalize well to other \Scratch applications or games developed in different programming languages.

\vspace{0.1cm}

\noindent \textbf{Internal Validity:} Randomized factors within the experiments pose another threat to validity since repeated executions of the same experiments will, by definition, lead to slightly different outcomes.
However, we expect 30 experiment repetitions for RQ1 and 100 test results for each project in RQ2 and RQ3 to suffice for assuming robust experiment results.
The threshold that determines at which surprise value a program is marked as a mutant was determined experimentally and may not work well for other programs.

\vspace{0.1cm}

\noindent \textbf{Construct Validity:} For evaluating the effectiveness of \Neatest in generating robust test suites, we used block coverage which is similar to statement coverage in text-based programming languages.
However, coverage is not always a good indicator because \Scratch programs may often already be covered by sending arbitrary inputs to the program.
For exactly this reason, we focused on selecting games that have a winning state and therefore pose a true challenge to test generators.
We answer whether networks can serve as test oracles by reporting the ratio of killed mutants.
However, not every mutant must necessarily represent erroneous behavior and may still be considered a valid program alternative by a human.

\subsection{RQ1: Can \Neatest Optimize Networks to Exercise \Scratch Games Reliably?}\label{sec:ResultsRQ1}



\begin{table}[t!]
    \caption{Reliable mean coverage, coverage effect size (A12), and number of reached winning states (Wins) of the random test generator (R) and \Neatest (N) during test generation. Boldface indicates strong statistical significance with $p$ <  0.05.}
    \label{tab:RQ1}
    \vspace{-1em}
    \resizebox{\columnwidth}{!}{
        \begin{tabular}{lrrrrrrr}
            \toprule
            \multicolumn{1}{r}{} & \multicolumn{4}{c}{Coverage} & \multicolumn{1}{r}{} & \multicolumn{2}{c}{Wins} \\
            \cline{2-5} \cline{7-8}
            Project            & R                                   & N                                 & A12                                        & $p$                                  & & R                                      & N                                    \\
            \midrule
            BirdShooter        & \GenBirdShooterMeanCovRandom        & \GenBirdShooterMeanCovNEAT        & \GenBirdShooterCoverageVDNonTrivial & \GenBirdShooterCoverageVDPVal & & \WinningStatesBirdShooterRandom & \WinningStatesBirdShooterNEAT  \\
            BrainGame          & \GenBrainGameMeanCovRandom          & \GenBrainGameMeanCovNEAT          & \GenBrainGameCoverageVDNonTrivial & \GenBrainGameCoverageVDPVal & & \WinningStatesBrainGameRandom & \WinningStatesBrainGameNEAT  \\
            CatchTheDots       & \GenCatchTheDotsMeanCovRandom       & \GenCatchTheDotsMeanCovNEAT       & \GenCatchTheDotsCoverageVDNonTrivial & \GenCatchTheDotsCoverageVDPVal & & \WinningStatesCatchTheDotsRandom & \WinningStatesCatchTheDotsNEAT  \\
            CatchTheGifts      & \GenCatchTheGiftsMeanCovRandom      & \GenCatchTheGiftsMeanCovNEAT      & \GenCatchTheGiftsCoverageVDNonTrivial & \GenCatchTheGiftsCoverageVDPVal & & \WinningStatesCatchTheGiftsRandom & \WinningStatesCatchTheGiftsNEAT  \\
            CatchingApples     & \GenCatchingApplesMeanCovRandom     & \GenCatchingApplesMeanCovNEAT     & \GenCatchingApplesCoverageVDNonTrivial & \GenCatchingApplesCoverageVDPVal & & \WinningStatesCatchingApplesRandom & \WinningStatesCatchingApplesNEAT  \\
            CityDefender       & \GenCityDefenderMeanCovRandom       & \GenCityDefenderMeanCovNEAT       & \GenCityDefenderCoverageVDNonTrivial & \GenCityDefenderCoverageVDPVal & & \WinningStatesCityDefenderRandom & \WinningStatesCityDefenderNEAT  \\
            DessertRally       & \GenDessertRallyMeanCovRandom       & \GenDessertRallyMeanCovNEAT       & \GenDessertRallyCoverageVDNonTrivial & \GenDessertRallyCoverageVDPVal & & \WinningStatesDessertRallyRandom & \WinningStatesDessertRallyNEAT  \\
            DieZauberlehrlinge & \GenDieZauberlehrlingeMeanCovRandom & \GenDieZauberlehrlingeMeanCovNEAT & \GenDieZauberlehrlingeCoverageVDNonTrivial & \GenDieZauberlehrlingeCoverageVDPVal & & \WinningStatesDieZauberlehrlingeRandom & \WinningStatesDieZauberlehrlingeNEAT  \\
            Dodgeball          & \GenDodgeballMeanCovRandom          & \GenDodgeballMeanCovNEAT          & \GenDodgeballCoverageVDNonTrivial & \GenDodgeballCoverageVDPVal & & \WinningStatesDodgeballRandom & \WinningStatesDodgeballNEAT  \\
            Dragons            & \GenDragonsMeanCovRandom            & \GenDragonsMeanCovNEAT            & \GenDragonsCoverageVDNonTrivial & \GenDragonsCoverageVDPVal & & \WinningStatesDragonsRandom & \WinningStatesDragonsNEAT  \\
            EndlessRunner      & \GenEndlessRunnerMeanCovRandom      & \GenEndlessRunnerMeanCovNEAT      & \GenEndlessRunnerCoverageVDNonTrivial & \GenEndlessRunnerCoverageVDPVal & & \WinningStatesEndlessRunnerRandom & \WinningStatesEndlessRunnerNEAT  \\
            FallingStars       & \GenFallingStarsMeanCovRandom       & \GenFallingStarsMeanCovNEAT       & \GenFallingStarsCoverageVDNonTrivial & \GenFallingStarsCoverageVDPVal & & \WinningStatesFallingStarsRandom & \WinningStatesFallingStarsNEAT  \\
            FinalFight         & \GenFinalFightMeanCovRandom         & \GenFinalFightMeanCovNEAT         & \GenFinalFightCoverageVDNonTrivial & \GenFinalFightCoverageVDPVal & & \WinningStatesFinalFightRandom & \WinningStatesFinalFightNEAT  \\
            FlappyParrot       & \GenFlappyParrotMeanCovRandom       & \GenFlappyParrotMeanCovNEAT       & \GenFlappyParrotCoverageVDNonTrivial & \GenFlappyParrotCoverageVDPVal & & \WinningStatesFlappyParrotRandom & \WinningStatesFlappyParrotNEAT  \\
            Frogger            & \GenFroggerMeanCovRandom            & \GenFroggerMeanCovNEAT            & \GenFroggerCoverageVDNonTrivial & \GenFroggerCoverageVDPVal & & \WinningStatesFroggerRandom & \WinningStatesFroggerNEAT  \\
            FruitCatching      & \GenFruitCatchingMeanCovRandom      & \GenFruitCatchingMeanCovNEAT      & \GenFruitCatchingCoverageVDNonTrivial & \GenFruitCatchingCoverageVDPVal & & \WinningStatesFruitCatchingRandom & \WinningStatesFruitCatchingNEAT  \\
            HackAttack         & \GenHackAttackMeanCovRandom         & \GenHackAttackMeanCovNEAT         & \GenHackAttackCoverageVDNonTrivial & \GenHackAttackCoverageVDPVal & & \WinningStatesHackAttackRandom & \WinningStatesHackAttackNEAT  \\
            LineUp             & \GenLineUpMeanCovRandom             & \GenLineUpMeanCovNEAT             & \GenLineUpCoverageVDNonTrivial & \GenLineUpCoverageVDPVal & & \WinningStatesLineUpRandom & \WinningStatesLineUpNEAT  \\
            OceanCleanup       & \GenOceanCleanupMeanCovRandom       & \GenOceanCleanupMeanCovNEAT       & \GenOceanCleanupCoverageVDNonTrivial & \GenOceanCleanupCoverageVDPVal & & \WinningStatesOceanCleanupRandom & \WinningStatesOceanCleanupNEAT  \\
            Pong               & \GenPongMeanCovRandom               & \GenPongMeanCovNEAT               & \GenPongCoverageVDNonTrivial               & \GenPongCoverageVDPVal               & & \WinningStatesPongRandom & \WinningStatesPongNEAT  \\
            RioShootout        & \GenRioShootoutMeanCovRandom        & \GenRioShootoutMeanCovNEAT        & \GenRioShootoutCoverageVDNonTrivial & \GenRioShootoutCoverageVDPVal & & \WinningStatesRioShootoutRandom & \WinningStatesRioShootoutNEAT  \\
            Snake              & \GenSnakeMeanCovRandom              & \GenSnakeMeanCovNEAT              & \GenSnakeCoverageVDNonTrivial              & \GenSnakeCoverageVDPVal              & & \WinningStatesSnakeRandom & \WinningStatesSnakeNEAT  \\
            SnowballFight      & \GenSnowballFightMeanCovRandom      & \GenSnowballFightMeanCovNEAT      & \GenSnowballFightCoverageVDNonTrivial & \GenSnowballFightCoverageVDPVal & & \WinningStatesSnowballFightRandom & \WinningStatesSnowballFightNEAT  \\
            SpaceOdyssey       & \GenSpaceOdysseyMeanCovRandom       & \GenSpaceOdysseyMeanCovNEAT       & \GenSpaceOdysseyCoverageVDNonTrivial & \GenSpaceOdysseyCoverageVDPVal & & \WinningStatesSpaceOdysseyRandom & \WinningStatesSpaceOdysseyNEAT  \\
            WhackAMole         & \GenWhackAMoleMeanCovRandom         & \GenWhackAMoleMeanCovNEAT         & \GenWhackAMoleCoverageVDNonTrivial & \GenWhackAMoleCoverageVDPVal & & \WinningStatesWhackAMoleRandom & \WinningStatesWhackAMoleNEAT  \\
            \midrule
            Mean               & \GenNonTrivialPMeanCovRandom        & \GenNonTrivialPMeanCovNEAT        & \GenMeanCoverageVDNonTrivial & - & & \WinningStatesAverageRandom & \WinningStatesAverageNEAT \\
            \bottomrule
        \end{tabular}
    }
\end{table}

Our first research question evaluates whether neural networks can be trained to exercise randomized \Scratch programs reliably. 
\cref{tab:RQ1} summarizes the experiment results and shows that the random tester and \Neatest reach generally high coverage values of 89.07\% and 94.60\%. This is because the programs tend to be small in
terms of their number of statements. However, \Scratch programs consist of many
domain-specific blocks specifically designed for game-behavior, such that it is
possible to implement fully functional game-behavior with very few blocks,
which in other programming languages might require hundreds of lines of code.
Blindly sending inputs, as random testing does, leads to the execution of a large
share of these blocks without, however, actually playing the game. This can be
seen when considering specifically how often an actual winning state was
reached. For example, in \emph{FallingStars} both approaches achieve almost the same average coverage (97.80\% and 100\%), although only \Neatest is able to cover statements related to the challenging winning state.
\cref{tab:RQ1} shows that \Neatest wins the game, on average, 20 times, while the random test generator has trouble meaningfully exercising the games and reaches the winning state only five times per game.
For three games, \Neatest fails to reach the winning state because either the allotted play duration is too short (\emph{DessertRally}), the game is too complex to evolve sufficiently optimized networks using the experiment parameters (\emph{Snake}), or more sophisticated evolution strategies are required to solve advanced search-problems, such as the maze problem~\cite{MazeNEAT}~(\emph{Dodgeball}).


\emph{CityDefender} is the only game for which \Neatest achieves less coverage than the random tester.
The statements responsible for these results are linked to relatively easy program states that the random tester can easily reach through randomly firing inputs at the program under test.
While \Neatest also sometimes performs the required actions, the networks must be specifically trained to send those inputs regardless of which randomized program state they are confronted with. However, the relatively long duration of a single playthrough in this game sometimes leads
to experiment repetitions in which the respective statements never get targeted.
The three reached winning states in \emph{CityDefender} that correspond to experiment repetitions in which a winning statement was actually targeted demonstrates that using a more sophisticated statement selection technique would allow \Neatest to train robust networks that reliably perform the required actions.

\begin{figure}[!tbp]
    \centering
    \includegraphics[width=\columnwidth]{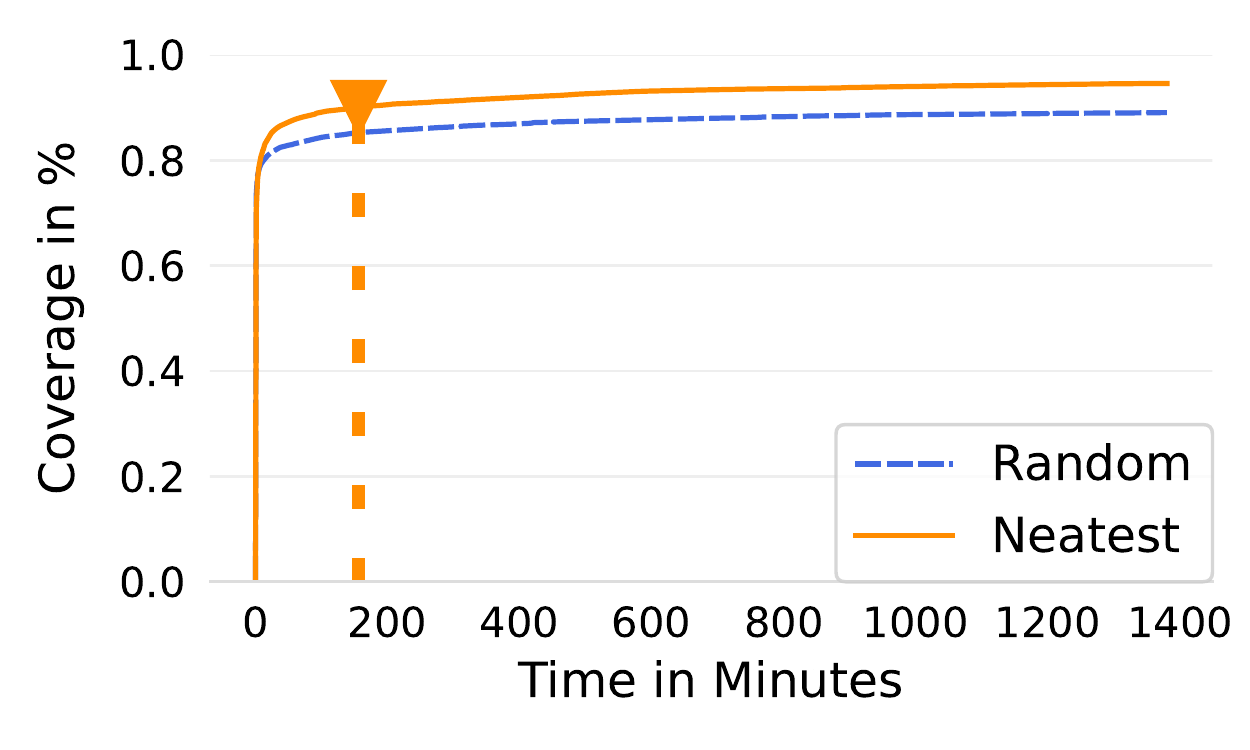}
    \caption{Coverage over time across all experiment repetitions and dataset programs with $\blacktriangledown$ representing an achieved total coverage value of 90\%.}
    \label{fig:CovOverTime}
\end{figure}

Having execution times of 23 hours may be recognized as a limiting factor of the proposed approach.
However, we selected this relatively high search budget in order to sufficiently explore those programs in the dataset that have exceptionally long fitness evaluation durations.
\cref{fig:CovOverTime} illustrates the achieved \emph{reliable} coverage over time for both approaches across all evaluation subjects.
The results show that \Neatest is more effective than random testing by covering more program statements at any stage of the search process.
Furthermore, the proposed approach covers 90\% of all program statements in less than three hours (156 min), while the random tester is not able to reach the same level of coverage at all.
Moreover, \Neatest fully covers \FullCoverageBlowOneHourNEAT/25 projects in all 30 repetitions within the first hour, indicating that the required search time is highly dependent on the complexity of the game.
In order to reduce the required search time for the more challenging programs, advanced search strategies, such as combining \Neatest with backpropagation \cite{NEAT-Backprop}, could be interesting avenues for future work.

\vspace{-1em}
\begin{figure}[H]
\includegraphics{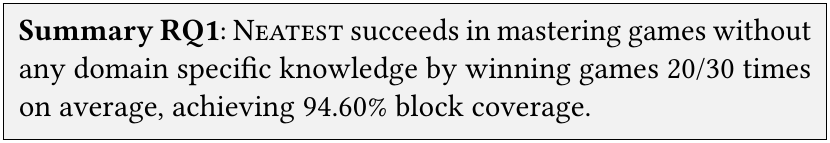}
\end{figure}
\subsection{RQ2: Are Dynamic Test Suites Robust Against Randomized Program Behavior?}

\begin{table}[t]
    \caption{Mean coverage and effect size (A12) during the test execution phase of dynamic (D) and static (S) test suites. The column \emph{Difference} shows coverage differences between the generation and execution phase. Values in boldface indicate strong statistical significance with $p$ <  0.05.}
    \label{tab:RQ2}
            \vspace{-1em}

    \resizebox{\columnwidth}{!}{
        \begin{tabular}{lrrrrrrrr}
            \toprule
            \multicolumn{1}{r}{} & \multicolumn{4}{c}{Coverage} & \multicolumn{1}{r}{} & \multicolumn{2}{c}{Difference} \\
            \cline{2-5} \cline{7-8}
            Project            & S                                    & D                                     & A12                                         & $p$                                   & & S                                          & D                                           \\
            \midrule
            BirdShooter        & \ExecBirdShooterMeanCovStatic        & \ExecBirdShooterMeanCovDynamic        & \ExecBirdShooterCoverageVDNonTrivial & \ExecBirdShooterCoverageVDPVal & & \ExecDifferenceStaticCovBirdShooter & \ExecDifferenceDynamicCovBirdShooter \\
            BrainGame          & \ExecBrainGameMeanCovStatic          & \ExecBrainGameMeanCovDynamic          & \ExecBrainGameCoverageVDNonTrivial & \ExecBrainGameCoverageVDPVal & & \ExecDifferenceStaticCovBrainGame & \ExecDifferenceDynamicCovBrainGame \\
            CatchTheDots       & \ExecCatchTheDotsMeanCovStatic       & \ExecCatchTheDotsMeanCovDynamic       & \ExecCatchTheDotsCoverageVDNonTrivial & \ExecCatchTheDotsCoverageVDPVal & & \ExecDifferenceStaticCovCatchTheDots & \ExecDifferenceDynamicCovCatchTheDots \\
            CatchTheGifts      & \ExecCatchTheGiftsMeanCovStatic      & \ExecCatchTheGiftsMeanCovDynamic      & \ExecCatchTheGiftsCoverageVDNonTrivial & \ExecCatchTheGiftsCoverageVDPVal & & \ExecDifferenceStaticCovCatchTheGifts & \ExecDifferenceDynamicCovCatchTheGifts \\
            CatchingApples     & \ExecCatchingApplesMeanCovStatic     & \ExecCatchingApplesMeanCovDynamic & \ExecCatchingApplesCoverageVDNonTrivial & \ExecCatchingApplesCoverageVDPVal & & \ExecDifferenceStaticCovCatchingApples & \ExecDifferenceDynamicCovCatchingApples \\
            CityDefender       & \ExecCityDefenderMeanCovStatic       & \ExecCityDefenderMeanCovDynamic       & \ExecCityDefenderCoverageVDNonTrivial & \ExecCityDefenderCoverageVDPVal & & \ExecDifferenceStaticCovCityDefender & \ExecDifferenceDynamicCovCityDefender \\
            DessertRally       & \ExecDessertRallyMeanCovStatic       & \ExecDessertRallyMeanCovDynamic       & \ExecDessertRallyCoverageVDNonTrivial & \ExecDessertRallyCoverageVDPVal & & \ExecDifferenceStaticCovDessertRally & \ExecDifferenceDynamicCovDessertRally \\
            DieZauberlehrlinge & \ExecDieZauberlehrlingeMeanCovStatic & \ExecDieZauberlehrlingeMeanCovDynamic & \ExecDieZauberlehrlingeCoverageVDNonTrivial & \ExecDieZauberlehrlingeCoverageVDPVal & & \ExecDifferenceStaticCovDieZauberlehrlinge & \ExecDifferenceDynamicCovDieZauberlehrlinge \\
            Dodgeball          & \ExecDodgeballMeanCovStatic          & \ExecDodgeballMeanCovDynamic          & \ExecDodgeballCoverageVDNonTrivial & \ExecDodgeballCoverageVDPVal & & \ExecDifferenceStaticCovDodgeball & \ExecDifferenceDynamicCovDodgeball \\
            Dragons            & \ExecDragonsMeanCovStatic            & \ExecDragonsMeanCovDynamic            & \ExecDragonsCoverageVDNonTrivial & \ExecDragonsCoverageVDPVal & & \ExecDifferenceStaticCovDragons & \ExecDifferenceDynamicCovDragons \\
            EndlessRunner      & \ExecEndlessRunnerMeanCovStatic      & \ExecEndlessRunnerMeanCovDynamic      & \ExecEndlessRunnerCoverageVDNonTrivial & \ExecEndlessRunnerCoverageVDPVal & & \ExecDifferenceStaticCovEndlessRunner & \ExecDifferenceDynamicCovEndlessRunner \\
            FallingStars       & \ExecFallingStarsMeanCovStatic       & \ExecFallingStarsMeanCovDynamic       & \ExecFallingStarsCoverageVDNonTrivial & \ExecFallingStarsCoverageVDPVal & & \ExecDifferenceStaticCovFallingStars & \ExecDifferenceDynamicCovFallingStars \\
            FinalFight         & \ExecFinalFightMeanCovStatic         & \ExecFinalFightMeanCovDynamic         & \ExecFinalFightCoverageVDNonTrivial & \ExecFinalFightCoverageVDPVal & & \ExecDifferenceStaticCovFinalFight & \ExecDifferenceDynamicCovFinalFight \\
            FlappyParrot       & \ExecFlappyParrotMeanCovStatic       & \ExecFlappyParrotMeanCovDynamic       & \ExecFlappyParrotCoverageVDNonTrivial & \ExecFlappyParrotCoverageVDPVal & & \ExecDifferenceStaticCovFlappyParrot & \ExecDifferenceDynamicCovFlappyParrot \\
            Frogger            & \ExecFroggerMeanCovStatic            & \ExecFroggerMeanCovDynamic            & \ExecFroggerCoverageVDNonTrivial & \ExecFroggerCoverageVDPVal & & \ExecDifferenceStaticCovFrogger & \ExecDifferenceDynamicCovFrogger \\
            FruitCatching      & \ExecFruitCatchingMeanCovStatic      & \ExecFruitCatchingMeanCovDynamic      & \ExecFruitCatchingCoverageVDNonTrivial & \ExecFruitCatchingCoverageVDPVal & & \ExecDifferenceStaticCovFruitCatching & \ExecDifferenceDynamicCovFruitCatching \\
            HackAttack         & \ExecHackAttackMeanCovStatic         & \ExecHackAttackMeanCovDynamic         & \ExecHackAttackCoverageVDNonTrivial & \ExecHackAttackCoverageVDPVal & & \ExecDifferenceStaticCovHackAttack & \ExecDifferenceDynamicCovHackAttack \\
            LineUp             & \ExecLineUpMeanCovStatic             & \ExecLineUpMeanCovDynamic             & \ExecLineUpCoverageVDNonTrivial & \ExecLineUpCoverageVDPVal & & \ExecDifferenceStaticCovLineUp & \ExecDifferenceDynamicCovLineUp \\
            OceanCleanup       & \ExecOceanCleanupMeanCovStatic       & \ExecOceanCleanupMeanCovDynamic       & \ExecOceanCleanupCoverageVDNonTrivial & \ExecOceanCleanupCoverageVDPVal & & \ExecDifferenceStaticCovOceanCleanup & \ExecDifferenceDynamicCovOceanCleanup \\
            Pong               & \ExecPongMeanCovStatic               & \ExecPongMeanCovDynamic               & \ExecPongCoverageVDNonTrivial & \ExecPongCoverageVDPVal & & \ExecDifferenceStaticCovPong & \ExecDifferenceDynamicCovPong \\
            RioShootout        & \ExecRioShootoutMeanCovStatic        & \ExecRioShootoutMeanCovDynamic        & \ExecRioShootoutCoverageVDNonTrivial & \ExecRioShootoutCoverageVDPVal & & \ExecDifferenceStaticCovRioShootout & \ExecDifferenceDynamicCovRioShootout \\
            Snake              & \ExecSnakeMeanCovStatic              & \ExecSnakeMeanCovDynamic              & \ExecSnakeCoverageVDNonTrivial & \ExecSnakeCoverageVDPVal & & \ExecDifferenceStaticCovSnake & \ExecDifferenceDynamicCovSnake \\
            \midrule
            Mean               & \ExecNonTrivialPMeanCovStatic        & \ExecNonTrivialPMeanCovDynamic        & \ExecMeanCoverageVDNonTrivial & - & & \ExecNonTrivialPDifferenceCovStatic & +\ExecNonTrivialPDifferenceCovDynamic \\
            \bottomrule
        \end{tabular}
    }
\end{table}

In order to evaluate the importance of training networks to \emph{reliably}
cover statements, RQ2 extracts static event sequences from the dynamic tests
generated for RQ1, and compares them in terms of the coverage variation
observed on executions with different random seeds. Each static test is the
result of executing a dynamic test on the program under test once. Hence, for the
particular seed of the random number generator on which the static test is derived, the coverage between both test types is by
construction identical. \cref{tab:RQ2} summarizes the effects of re-running the
resulting tests for ten different random seeds on the respective programs. The
results show that dynamic tests are robust against randomized
programs---coverage even increases slightly by 1.22\% compared to the test
generation phase. In contrast, static tests have a drop of 8.17\% in
coverage, which is substantial considering the generally high coverage observed
in RQ1.

The slight increase in coverage for the dynamic test suites may seem
surprising, but can be explained by the way coverage is measured: In contrast
to RQ1, which measured how many statements were \emph{reliably} reached by a test
generator, here we count how many statements are covered on average over the
100 test executions. Thus, also statements that are only covered in some of the
runs contribute to the overall coverage. The drop of 8.17\% coverage for static tests
demonstrates the need for robust test suites when testing randomized programs.
This clear advantage of dynamic tests becomes even more apparent when looking
at the \atwelve values (on average 0.82), which indicate that for every program in the dataset,
networks perform significantly better than static test sequences.

\vspace{-1em}
\begin{figure}[H]
\includegraphics{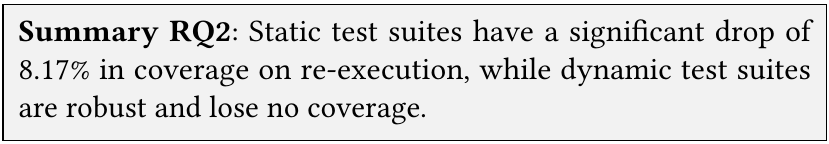}
\end{figure}


\subsection{RQ3: Can Networks Serve as Test Oracles?}

The randomized behavior of games impedes the generation of reliable test oracles.
Hence, RQ3 aims to investigate if the networks of dynamic tests can
serve as test oracles that can distinguish randomized program scenarios from
faulty behavior. We generated \GeneratedMutants mutants using
eight different mutation operators across all games in our dataset and checked
whether the networks were able to identify them. 
\Cref{tab:Mutants} demonstrates that the networks managed to achieve an average mutation score of \MutationScore\% per project, while maintaining a relatively low false-positive rate of \FalsePositiveRate\%.

\begin{figure}[!tbp]
    \centering
    \includegraphics[width=\columnwidth]{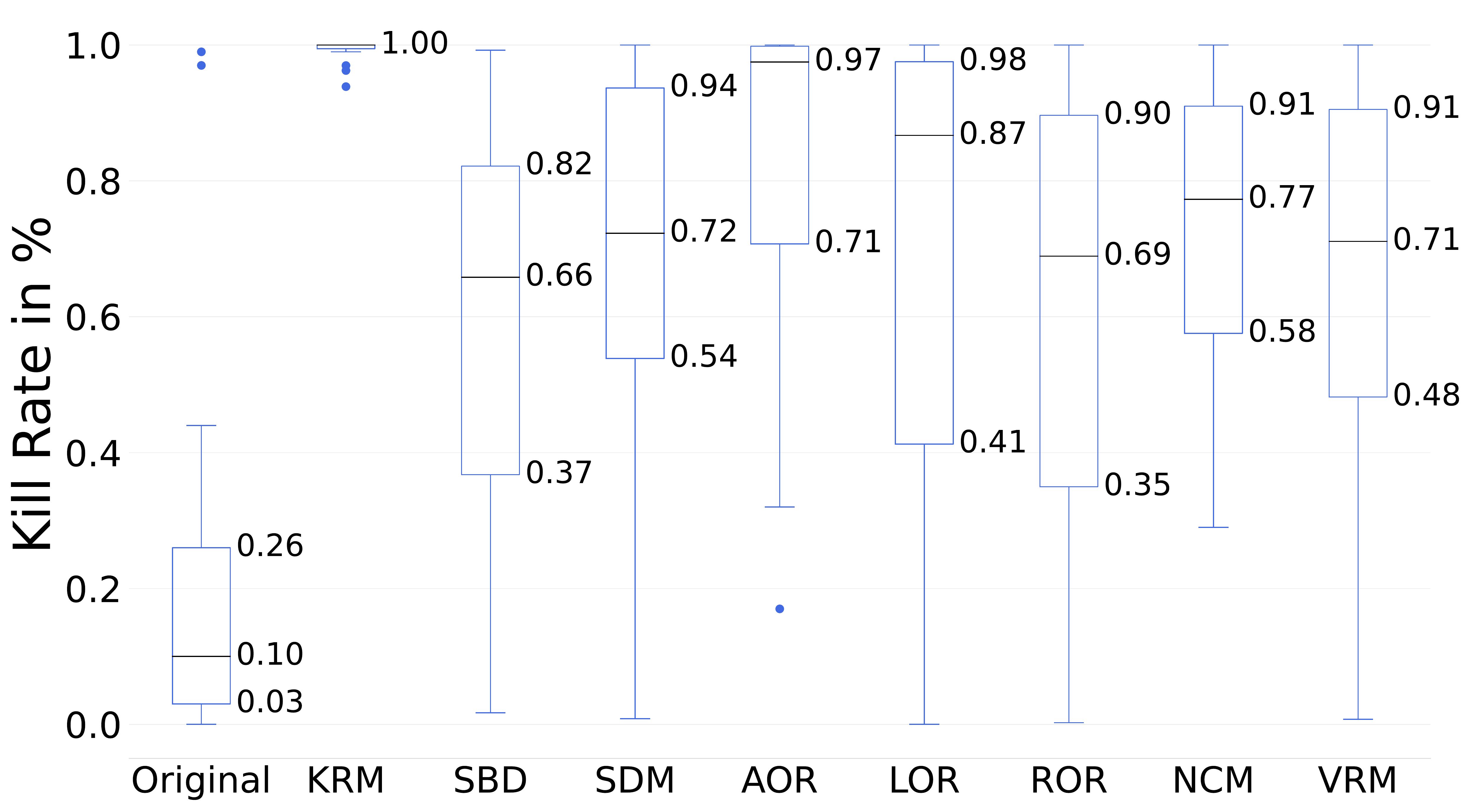}
    \vspace{-2em}
    \caption{Mutant kill rates across all mutation operators.}
    \label{fig:RQ3-CategoryKillRate}
    \vspace{-1em}
\end{figure}

\Cref{fig:RQ3-CategoryKillRate} indicates that the average false-positive rate
is deceiving as there are two extreme outliers that have a strong influence on the mean, while the median is considerably lower with a value of only 10\%. In those two programs, namely
\emph{Dragons} and \emph{FinalFight}, the networks are
confronted with many sprites having attributes that are distinct in almost
every program execution, resulting in activation
traces that are entirely different from previously observed ones, and thus seem
suspicious. This problem could be solved by either increasing the number of
recorded ground truth traces or through a more sophisticated abstraction layer
by restricting the set of observed hidden nodes to nodes that reside in hidden
layers that are further away from the input layer.

\Neatest detects the least mutants for \emph{CityDefender}, which is a direct consequence of the poorly trained networks (see \cref{sec:ResultsRQ1}).
The lowest mutation kill rate can be observed in
\cref{fig:RQ3-CategoryKillRate} for the \emph{Single Block Deletion} mutation
operator, because \Scratch programs often contain statements that do not change
the program behavior but instead offer purely visual effects.
Depending on the scope of what is recognized as faulty behavior, these visual
deviations could be detected by collecting more fine-grained input features,
such as the visual effects applied to the individual sprites.
The highest mutation kill rate was achieved for the \emph{Key Replacement Mutation} operator
(\CategoryKRMKillRate\%), in which an event handler's observed key gets replaced
with a randomly selected one. In many of these cases, the mutant is detected
due to the network's dynamic adaption to changes in the input and output
space~(cf. \cref{sec:FeatureExtraction}).

\begin{table}[tb]
    \centering
    \caption{Generated (capped at 50 mutants per operator and game), killed and false-positive (FP) marked mutants.}
    \label{tab:Mutants}
        \vspace{-1em}
    \resizebox{\columnwidth}{!}{
        \begin{tabular}{lrrrlrrr}
            \toprule
            Project & \rotatebox{90}{\texttt{\#} Generated} & \rotatebox{90}{Killed \%} & \rotatebox{90}{FP \%} &
            Project & \rotatebox{90}{\texttt{\#} Generated} & \rotatebox{90}{Killed \%} & \rotatebox{90}{FP \%} \\
            \midrule
            BirdShooter        & \BirdShooterMutants        & \BirdShooterKillRate        & \BirdShooterFalsePositivesRate          & FlappyParrot  & \FlappyParrotMutants & \FlappyParrotKillRate & \FlappyParrotFalsePositivesRate \\
            BrainGame          & \BrainGameMutants          & \BrainGameKillRate          & \BrainGameFalsePositivesRate          & Frogger       & \FroggerMutants & \FroggerKillRate & \FroggerFalsePositivesRate \\
            CatchTheDots       & \CatchTheDotsMutants       & \CatchTheDotsKillRate       & \CatchTheDotsFalsePositivesRate         & FruitCatching & \FruitCatchingMutants & \FruitCatchingKillRate & \FruitCatchingFalsePositivesRate \\
            CatchTheGifts      & \CatchTheGiftsMutants      & \CatchTheGiftsKillRate      & \CatchTheGiftsFalsePositivesRate        & HackAttack    & \HackAttackMutants & \HackAttackKillRate & \HackAttackFalsePositivesRate \\
            CatchingApples     & \CatchingApplesMutants     & \CatchingApplesKillRate     & \CatchingApplesFalsePositivesRate       & LineUp        & \LineUpMutants & \LineUpKillRate & \LineUpFalsePositivesRate \\
            CityDefender       & \CityDefenderMutants       & \CityDefenderKillRate       & \CityDefenderFalsePositivesRate         & OceanCleanup  & \OceanCleanupMutants & \OceanCleanupKillRate & \OceanCleanupFalsePositivesRate \\
            DessertRally       & \DessertRallyMutants       & \DessertRallyKillRate       & \DessertRallyFalsePositivesRate         & Pong          & \PongMutants & \PongKillRate & \PongFalsePositivesRate \\
            DieZauberlehrlinge & \DieZauberlehrlingeMutants & \DieZauberlehrlingeKillRate & \DieZauberlehrlingeFalsePositivesRate   & RioShootout   & \RioShootoutMutants & \RioShootoutKillRate & \RioShootoutFalsePositivesRate \\
            Dodgeball          & \DodgeballMutants          & \DodgeballKillRate          & \DodgeballFalsePositivesRate          & Snake         & \SnakeMutants & \SnakeKillRate & \SnakeFalsePositivesRate \\
            Dragons            & \DragonsMutants            & \DragonsKillRate            & \DragonsFalsePositivesRate            & SnowballFight & \SnowballFightMutants & \SnowballFightKillRate & \SnowballFightFalsePositivesRate \\
            EndlessRunner      & \EndlessRunnerMutants      & \EndlessRunnerKillRate      & \EndlessRunnerFalsePositivesRate        & SpaceOdyssey & \SpaceOdysseyMutants & \SpaceOdysseyKillRate & \SpaceOdysseyFalsePositivesRate \\
            FallingStars       & \FallingStarsMutants       & \FallingStarsKillRate       & \FallingStarsFalsePositivesRate        & WhackAMole    & \WhackAMoleMutants & \WhackAMoleKillRate & \WhackAMoleFalsePositivesRate \\
            FinalFight         & \FinalFightMutants         & \FinalFightKillRate         & \FinalFightFalsePositivesRate         & Mean          & \GeneratedMutantsMean & \MutationScore & \FalsePositiveRate \\
            \bottomrule
        \end{tabular}
    }
        \vspace{-1em}

\end{table}

\vspace{-1em}
 \begin{figure}[H]
\includegraphics{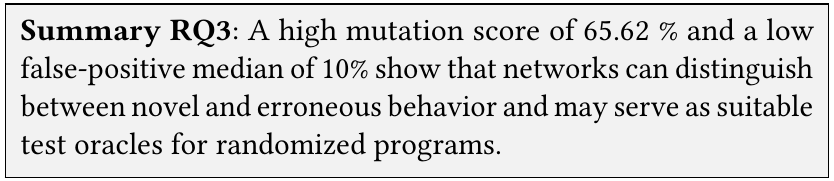}
\end{figure}

\section{Related Work}\label{RelatedWork}

Due to the increasing popularity of NEAT, many variations have been proposed \cite{HyperNEAT, LSTM-NEAT, NEAT-stochastic,
NEAT-Snap, FS-NEAT} and applied to master games in many different gaming
environments, such as Atari \cite{Neat-Atari} and real-time based games
\cite{NERO, Neat-RTS}. 
Since, to the best of our knowledge, no previous work exists that
uses neuroevolution for block-based programming environments like \Scratch, we
applied the established variant of NEAT, and will evaluate improvements in
future work.

Many reinforcement learning techniques have been shown to be capable
of learning to play different game genres such as strategy
games~\cite{RL-Strategy, RL-Strategy2}, arcade games \cite{RL-Arcade},
first-person shooter games~\cite{RL-FPS,RL-FPS2}, and casual games
\cite{RL-Casual}. We use neuroevolution rather than
alternative reinforcement learning techniques since NEAT allows easy adaptation
to changes in the input space. Furthermore, prior reinforcement learning-based
approaches reward an agent for achieving game-specific objectives. In contrast,
\Neatest explores the source code and automatically derives code-based objectives to
cover specific target statements, thus implicitly learning to play the games.

The \emph{Likelihood-based Surprise Adequacy} metric, which we use as a test
oracle, has previously been used for prioritizing test inputs for image
classification tasks \cite{SA}. Kim et al. \cite{SAI} reduced the cost of
calculating the \emph{Surprise Adequacy} by replacing the kernel density
estimation with the Mahalanobis distance \cite{Mahalanobis}. In future
research, we will explore other metrics, such as the certainty of networks in their predictions \cite{DeepGini}, and their usability as test oracles.

Nowadays, games are either tested manually \cite{politowski2021survey, ferre2009playability, diah2010usability} or with semi-automated approaches \cite{schaefer2013crushinator, cho2010online, smith2009computational}.
Furthermore, many approaches do not aim to detect bugs but rather to evaluate the usability of games~\cite{RL-Playtesting, desurvire2004using, korhonen2006playability}.
While recently  automated game testing has been explored~\cite{zheng2019wuji, ariyurek2019automated, ICARUS, politowski2022towards}, overall the research area is still young, and to the best of our knowledge no automated approaches exist that generate tests using neuroevolution.

\section{Conclusions}\label{Conclusions}

Games pose a considerable challenge for automated test generation because they
require the test generator to master a given game. Furthermore, games tend to be heavily randomized to keep players
engaged, which prevents traditional static test suites from testing such
programs reliably, and static test assertions from differentiating between novel and erroneous program
states. In an attempt to solve these challenges, this work combines 
search-based software testing with neuroevolution to produce
a novel type of test suites: Dynamic test suites consist of neural networks
trained via neuroevolution to reach program states such as the winning state
reliably, regardless of the encountered program scenarios. By observing the
network structure and measuring how surprised the networks are by a given
program state,  the networks serve as test
oracles that detect incorrect program behavior.
Our experiments confirm that \Neatest manages to master and test 
\Scratch games. 

The \Neatest approach provides an initial foundation for combining automated
testing with neuroevolution. The \Scratch games we considered in our prototype
are often similar to classic arcade games, a common target for neuroevolution
research. In principle, \Neatest is also applicable to other domains, such as
Android or iOS games, but adaptation will require engineering effort for
feature extraction and implementing the play loop. Integrating recent advances
in neuroevolution, such as deep learning extensions of NEAT \cite{DeepNEAT}, will
be of interest in order to generalize further beyond arcade games. Furthermore,
it may be worth exploring if \Neatest provides a way to overcome the problems
caused by randomized program behavior outside the domain of games (i.e., flaky
tests). Finally, 
the Surprise
Adequacy metric 
originates from a highly active area of research, and further research with new
and alternative metrics may further improve the detection rates for erroneous
program behavior. \Neatest and the hosting \Whisker testing framework are publicly available at:
	 \url{https://github.com/se2p/whisker}

\begin{acks}
  This work is supported by \grantsponsor{FR 2955/3-1}{DFG project
    FR2955/3-1 “TENDER-BLOCK: Testing, Debugging, and Repairing
    Blocks-based
    Programs”}{https://gepris.dfg.de/gepris/projekt/418126274} and the
  Federal Ministry of Education and Research through project 01JA2021
  (primary::programming) as part of the ``Qualitätsoffensive
  Lehrerbildung'', a joint initiative of the Federal Government and
  the Länder. The authors are responsible for this publication's
  content.

\end{acks}

\balance

\bibliographystyle{ACM-Reference-Format}
  \bibliography{related}
\end{document}

%% file: macros.tex
\newcommand{\GenBirdShooterMeanCovRandom}{98.60\xspace}

\newcommand{\GenBirdShooterMeanCovNEAT}{100\xspace}

\newcommand{\GenBrainGameMeanCovRandom}{89.34\xspace}

\newcommand{\GenBrainGameMeanCovNEAT}{100\xspace}

\newcommand{\GenCatchTheDotsMeanCovRandom}{97.64\xspace}

\newcommand{\GenCatchTheDotsMeanCovNEAT}{100\xspace}

\newcommand{\GenCatchTheGiftsMeanCovRandom}{89.80\xspace}

\newcommand{\GenCatchTheGiftsMeanCovNEAT}{100\xspace}

\newcommand{\GenCatchingApplesMeanCovRandom}{98.53\xspace}

\newcommand{\GenCatchingApplesMeanCovNEAT}{100\xspace}

\newcommand{\GenCityDefenderMeanCovRandom}{74.30\xspace}

\newcommand{\GenCityDefenderMeanCovNEAT}{70.62\xspace}

\newcommand{\GenDessertRallyMeanCovRandom}{93.58\xspace}

\newcommand{\GenDessertRallyMeanCovNEAT}{93.90\xspace}

\newcommand{\GenDieZauberlehrlingeMeanCovRandom}{75.50\xspace}

\newcommand{\GenDieZauberlehrlingeMeanCovNEAT}{92.17\xspace}

\newcommand{\GenDodgeballMeanCovRandom}{94.87\xspace}

\newcommand{\GenDodgeballMeanCovNEAT}{96.15\xspace}

\newcommand{\GenDragonsMeanCovRandom}{88.30\xspace}

\newcommand{\GenDragonsMeanCovNEAT}{89.01\xspace}

\newcommand{\GenEndlessRunnerMeanCovRandom}{73.95\xspace}

\newcommand{\GenEndlessRunnerMeanCovNEAT}{90.52\xspace}

\newcommand{\GenFallingStarsMeanCovRandom}{97.80\xspace}

\newcommand{\GenFallingStarsMeanCovNEAT}{100\xspace}

\newcommand{\GenFinalFightMeanCovRandom}{97.50\xspace}

\newcommand{\GenFinalFightMeanCovNEAT}{99.65\xspace}

\newcommand{\GenFlappyParrotMeanCovRandom}{92.61\xspace}

\newcommand{\GenFlappyParrotMeanCovNEAT}{100\xspace}

\newcommand{\GenFroggerMeanCovRandom}{88.69\xspace}

\newcommand{\GenFroggerMeanCovNEAT}{95.74\xspace}

\newcommand{\GenFruitCatchingMeanCovRandom}{69.09\xspace}

\newcommand{\GenFruitCatchingMeanCovNEAT}{100\xspace}

\newcommand{\GenHackAttackMeanCovRandom}{86.02\xspace}

\newcommand{\GenHackAttackMeanCovNEAT}{97.38\xspace}

\newcommand{\GenLineUpMeanCovRandom}{96.00\xspace}

\newcommand{\GenLineUpMeanCovNEAT}{100\xspace}

\newcommand{\GenOceanCleanupMeanCovRandom}{73.65\xspace}

\newcommand{\GenOceanCleanupMeanCovNEAT}{78.61\xspace}

\newcommand{\GenPongMeanCovRandom}{94.22\xspace}

\newcommand{\GenPongMeanCovNEAT}{100\xspace}

\newcommand{\GenRioShootoutMeanCovRandom}{94.40\xspace}

\newcommand{\GenRioShootoutMeanCovNEAT}{94.40\xspace}

\newcommand{\GenSnakeMeanCovRandom}{95.00\xspace}

\newcommand{\GenSnakeMeanCovNEAT}{95.00\xspace}

\newcommand{\GenSnowballFightMeanCovRandom}{94.87\xspace}

\newcommand{\GenSnowballFightMeanCovNEAT}{96.32\xspace}

\newcommand{\GenSpaceOdysseyMeanCovRandom}{92.70\xspace}

\newcommand{\GenSpaceOdysseyMeanCovNEAT}{95.17\xspace}

\newcommand{\GenWhackAMoleMeanCovRandom}{79.64\xspace}

\newcommand{\GenWhackAMoleMeanCovNEAT}{80.44\xspace}

\newcommand{\GenNonTrivialPMeanCovRandom}{89.07\xspace}
\newcommand{\GenNonTrivialPMeanCovNEAT}{94.60\xspace}

\newcommand{\GenBirdShooterCoverageVDPVal}{< 0.01\xspace}

\newcommand{\GenBrainGameCoverageVDPVal}{< 0.01\xspace}

\newcommand{\GenCatchTheDotsCoverageVDPVal}{< 0.01\xspace}

\newcommand{\GenCatchTheGiftsCoverageVDPVal}{< 0.01\xspace}

\newcommand{\GenCatchingApplesCoverageVDPVal}{< 0.01\xspace}

\newcommand{\GenCityDefenderCoverageVDPVal}{< 0.01\xspace}

\newcommand{\GenDessertRallyCoverageVDPVal}{0.02\xspace}

\newcommand{\GenDieZauberlehrlingeCoverageVDPVal}{< 0.01\xspace}

\newcommand{\GenDodgeballCoverageVDPVal}{< 0.01\xspace}

\newcommand{\GenDragonsCoverageVDPVal}{0.17\xspace}

\newcommand{\GenEndlessRunnerCoverageVDPVal}{< 0.01\xspace}

\newcommand{\GenFallingStarsCoverageVDPVal}{< 0.01\xspace}

\newcommand{\GenFinalFightCoverageVDPVal}{< 0.01\xspace}

\newcommand{\GenFlappyParrotCoverageVDPVal}{< 0.01\xspace}

\newcommand{\GenFroggerCoverageVDPVal}{< 0.01\xspace}

\newcommand{\GenFruitCatchingCoverageVDPVal}{< 0.01\xspace}

\newcommand{\GenHackAttackCoverageVDPVal}{< 0.01\xspace}

\newcommand{\GenLineUpCoverageVDPVal}{< 0.01\xspace}

\newcommand{\GenOceanCleanupCoverageVDPVal}{0.01\xspace}

\newcommand{\GenPongCoverageVDPVal}{< 0.01\xspace}

\newcommand{\GenRioShootoutCoverageVDPVal}{1.00\xspace}

\newcommand{\GenSnakeCoverageVDPVal}{1.00\xspace}

\newcommand{\GenSnowballFightCoverageVDPVal}{< 0.01\xspace}

\newcommand{\GenSpaceOdysseyCoverageVDPVal}{< 0.01\xspace}

\newcommand{\GenWhackAMoleCoverageVDPVal}{0.62\xspace}

\newcommand{\GenBirdShooterCoverageVDNonTrivial}{\textbf{0.98\xspace}}

\newcommand{\GenBrainGameCoverageVDNonTrivial}{\textbf{0.95\xspace}}

\newcommand{\GenCatchTheDotsCoverageVDNonTrivial}{\textbf{0.82\xspace}}

\newcommand{\GenCatchTheGiftsCoverageVDNonTrivial}{\textbf{0.67\xspace}}

\newcommand{\GenCatchingApplesCoverageVDNonTrivial}{\textbf{0.68\xspace}}

\newcommand{\GenCityDefenderCoverageVDNonTrivial}{\textbf{0.28\xspace}}

\newcommand{\GenDessertRallyCoverageVDNonTrivial}{\textbf{0.58\xspace}}

\newcommand{\GenDieZauberlehrlingeCoverageVDNonTrivial}{\textbf{1.00\xspace}}

\newcommand{\GenDodgeballCoverageVDNonTrivial}{\textbf{1.00\xspace}}

\newcommand{\GenDragonsCoverageVDNonTrivial}{0.61\xspace}

\newcommand{\GenEndlessRunnerCoverageVDNonTrivial}{\textbf{0.90\xspace}}

\newcommand{\GenFallingStarsCoverageVDNonTrivial}{\textbf{1.00\xspace}}

\newcommand{\GenFinalFightCoverageVDNonTrivial}{\textbf{1.00\xspace}}

\newcommand{\GenFlappyParrotCoverageVDNonTrivial}{\textbf{1.00\xspace}}

\newcommand{\GenFroggerCoverageVDNonTrivial}{\textbf{1.00\xspace}}

\newcommand{\GenFruitCatchingCoverageVDNonTrivial}{\textbf{1.00\xspace}}

\newcommand{\GenHackAttackCoverageVDNonTrivial}{\textbf{1.00\xspace}}

\newcommand{\GenLineUpCoverageVDNonTrivial}{\textbf{1.00\xspace}}

\newcommand{\GenOceanCleanupCoverageVDNonTrivial}{\textbf{0.69\xspace}}

\newcommand{\GenPongCoverageVDNonTrivial}{\textbf{0.72\xspace}}

\newcommand{\GenRioShootoutCoverageVDNonTrivial}{0.50\xspace}

\newcommand{\GenSnakeCoverageVDNonTrivial}{0.50\xspace}

\newcommand{\GenSnowballFightCoverageVDNonTrivial}{\textbf{0.78\xspace}}

\newcommand{\GenSpaceOdysseyCoverageVDNonTrivial}{\textbf{0.67\xspace}}

\newcommand{\GenWhackAMoleCoverageVDNonTrivial}{0.54\xspace}

\newcommand{\GenMeanCoverageVDNonTrivial}{0.80\xspace}

\newcommand{\ExecBirdShooterMeanCovStatic}{94.76\xspace}
\newcommand{\ExecBirdShooterMeanCovDynamic}{99.59\xspace}

\newcommand{\ExecDifferenceStaticCovBirdShooter}{-5.24\xspace}
\newcommand{\ExecDifferenceDynamicCovBirdShooter}{-0.41\xspace}

\newcommand{\ExecBrainGameMeanCovStatic}{23.53\xspace}
\newcommand{\ExecBrainGameMeanCovDynamic}{95.76\xspace}

\newcommand{\ExecDifferenceStaticCovBrainGame}{-76.47\xspace}
\newcommand{\ExecDifferenceDynamicCovBrainGame}{-4.24\xspace}

\newcommand{\ExecCatchTheDotsMeanCovStatic}{99.40\xspace}
\newcommand{\ExecCatchTheDotsMeanCovDynamic}{100\xspace}

\newcommand{\ExecDifferenceStaticCovCatchTheDots}{-0.60\xspace}
\newcommand{\ExecDifferenceDynamicCovCatchTheDots}{0.00\xspace}

\newcommand{\ExecCatchTheGiftsMeanCovStatic}{99.47\xspace}
\newcommand{\ExecCatchTheGiftsMeanCovDynamic}{100\xspace}

\newcommand{\ExecDifferenceStaticCovCatchTheGifts}{-0.53\xspace}
\newcommand{\ExecDifferenceDynamicCovCatchTheGifts}{0.00\xspace}

\newcommand{\ExecCatchingApplesMeanCovStatic}{96.17\xspace}
\newcommand{\ExecCatchingApplesMeanCovDynamic}{100\xspace}

\newcommand{\ExecDifferenceStaticCovCatchingApples}{-3.83\xspace}
\newcommand{\ExecDifferenceDynamicCovCatchingApples}{0.00\xspace}

\newcommand{\ExecCityDefenderMeanCovStatic}{72.91\xspace}
\newcommand{\ExecCityDefenderMeanCovDynamic}{81.02\xspace}

\newcommand{\ExecDifferenceStaticCovCityDefender}{2.29\xspace}
\newcommand{\ExecDifferenceDynamicCovCityDefender}{10.40\xspace}

\newcommand{\ExecDessertRallyMeanCovStatic}{93.65\xspace}
\newcommand{\ExecDessertRallyMeanCovDynamic}{93.97\xspace}

\newcommand{\ExecDifferenceStaticCovDessertRally}{-0.25\xspace}
\newcommand{\ExecDifferenceDynamicCovDessertRally}{0.07\xspace}

\newcommand{\ExecDieZauberlehrlingeMeanCovStatic}{78.66\xspace}
\newcommand{\ExecDieZauberlehrlingeMeanCovDynamic}{94.92\xspace}

\newcommand{\ExecDifferenceStaticCovDieZauberlehrlinge}{-13.51\xspace}
\newcommand{\ExecDifferenceDynamicCovDieZauberlehrlinge}{2.75\xspace}

\newcommand{\ExecDodgeballMeanCovStatic}{95.17\xspace}
\newcommand{\ExecDodgeballMeanCovDynamic}{95.83\xspace}

\newcommand{\ExecDifferenceStaticCovDodgeball}{-0.98\xspace}
\newcommand{\ExecDifferenceDynamicCovDodgeball}{-0.32\xspace}

\newcommand{\ExecDragonsMeanCovStatic}{90.95\xspace}
\newcommand{\ExecDragonsMeanCovDynamic}{91.02\xspace}

\newcommand{\ExecDifferenceStaticCovDragons}{1.94\xspace}
\newcommand{\ExecDifferenceDynamicCovDragons}{2.01\xspace}

\newcommand{\ExecEndlessRunnerMeanCovStatic}{84.65\xspace}
\newcommand{\ExecEndlessRunnerMeanCovDynamic}{94.92\xspace}

\newcommand{\ExecDifferenceStaticCovEndlessRunner}{-5.87\xspace}
\newcommand{\ExecDifferenceDynamicCovEndlessRunner}{4.40\xspace}

\newcommand{\ExecFallingStarsMeanCovStatic}{98.57\xspace}
\newcommand{\ExecFallingStarsMeanCovDynamic}{99.69\xspace}

\newcommand{\ExecDifferenceStaticCovFallingStars}{-1.43\xspace}
\newcommand{\ExecDifferenceDynamicCovFallingStars}{-0.31\xspace}

\newcommand{\ExecFinalFightMeanCovStatic}{96.07\xspace}
\newcommand{\ExecFinalFightMeanCovDynamic}{99.69\xspace}

\newcommand{\ExecDifferenceStaticCovFinalFight}{-3.58\xspace}
\newcommand{\ExecDifferenceDynamicCovFinalFight}{0.04\xspace}

\newcommand{\ExecFlappyParrotMeanCovStatic}{93.59\xspace}
\newcommand{\ExecFlappyParrotMeanCovDynamic}{96.41\xspace}

\newcommand{\ExecDifferenceStaticCovFlappyParrot}{-6.41\xspace}
\newcommand{\ExecDifferenceDynamicCovFlappyParrot}{-3.59\xspace}

\newcommand{\ExecFroggerMeanCovStatic}{93.95\xspace}
\newcommand{\ExecFroggerMeanCovDynamic}{95.72\xspace}

\newcommand{\ExecDifferenceStaticCovFrogger}{-1.79\xspace}
\newcommand{\ExecDifferenceDynamicCovFrogger}{-0.02\xspace}

\newcommand{\ExecFruitCatchingMeanCovStatic}{77.69\xspace}
\newcommand{\ExecFruitCatchingMeanCovDynamic}{96.70\xspace}

\newcommand{\ExecDifferenceStaticCovFruitCatching}{-22.31\xspace}
\newcommand{\ExecDifferenceDynamicCovFruitCatching}{-3.30\xspace}

\newcommand{\ExecHackAttackMeanCovStatic}{83.84\xspace}
\newcommand{\ExecHackAttackMeanCovDynamic}{92.06\xspace}

\newcommand{\ExecDifferenceStaticCovHackAttack}{-13.54\xspace}
\newcommand{\ExecDifferenceDynamicCovHackAttack}{-5.32\xspace}

\newcommand{\ExecLineUpMeanCovStatic}{71.03\xspace}
\newcommand{\ExecLineUpMeanCovDynamic}{99.46\xspace}

\newcommand{\ExecDifferenceStaticCovLineUp}{-28.97\xspace}
\newcommand{\ExecDifferenceDynamicCovLineUp}{-0.54\xspace}

\newcommand{\ExecOceanCleanupMeanCovStatic}{92.38\xspace}
\newcommand{\ExecOceanCleanupMeanCovDynamic}{98.51\xspace}

\newcommand{\ExecDifferenceStaticCovOceanCleanup}{13.77\xspace}
\newcommand{\ExecDifferenceDynamicCovOceanCleanup}{19.90\xspace}

\newcommand{\ExecPongMeanCovStatic}{89.93\xspace}
\newcommand{\ExecPongMeanCovDynamic}{98.00\xspace}

\newcommand{\ExecDifferenceStaticCovPong}{-10.07\xspace}
\newcommand{\ExecDifferenceDynamicCovPong}{-2.00\xspace}

\newcommand{\ExecRioShootoutMeanCovStatic}{85.10\xspace}
\newcommand{\ExecRioShootoutMeanCovDynamic}{96.67\xspace}

\newcommand{\ExecDifferenceStaticCovRioShootout}{-9.30\xspace}
\newcommand{\ExecDifferenceDynamicCovRioShootout}{2.27\xspace}

\newcommand{\ExecSnakeMeanCovStatic}{93.15\xspace}
\newcommand{\ExecSnakeMeanCovDynamic}{95.26\xspace}

\newcommand{\ExecDifferenceStaticCovSnake}{-1.85\xspace}
\newcommand{\ExecDifferenceDynamicCovSnake}{0.26\xspace}

\newcommand{\ExecNonTrivialPMeanCovStatic}{86.43\xspace}
\newcommand{\ExecNonTrivialPMeanCovDynamic}{95.83\xspace}

\newcommand{\ExecNonTrivialPDifferenceCovStatic}{-8.17\xspace}
\newcommand{\ExecNonTrivialPDifferenceCovDynamic}{1.22\xspace}

\newcommand{\ExecBirdShooterCoverageVDPVal}{< 0.01\xspace}

\newcommand{\ExecBrainGameCoverageVDPVal}{< 0.01\xspace}

\newcommand{\ExecCatchTheDotsCoverageVDPVal}{< 0.01\xspace}

\newcommand{\ExecCatchTheGiftsCoverageVDPVal}{< 0.01\xspace}

\newcommand{\ExecCatchingApplesCoverageVDPVal}{< 0.01\xspace}

\newcommand{\ExecCityDefenderCoverageVDPVal}{< 0.01\xspace}

\newcommand{\ExecDessertRallyCoverageVDPVal}{< 0.01\xspace}

\newcommand{\ExecDieZauberlehrlingeCoverageVDPVal}{< 0.01\xspace}

\newcommand{\ExecDodgeballCoverageVDPVal}{< 0.01\xspace}

\newcommand{\ExecDragonsCoverageVDPVal}{< 0.01\xspace}

\newcommand{\ExecEndlessRunnerCoverageVDPVal}{< 0.01\xspace}

\newcommand{\ExecFallingStarsCoverageVDPVal}{< 0.01\xspace}

\newcommand{\ExecFinalFightCoverageVDPVal}{< 0.01\xspace}

\newcommand{\ExecFlappyParrotCoverageVDPVal}{< 0.01\xspace}

\newcommand{\ExecFroggerCoverageVDPVal}{< 0.01\xspace}

\newcommand{\ExecFruitCatchingCoverageVDPVal}{< 0.01\xspace}

\newcommand{\ExecHackAttackCoverageVDPVal}{< 0.01\xspace}

\newcommand{\ExecLineUpCoverageVDPVal}{< 0.01\xspace}

\newcommand{\ExecOceanCleanupCoverageVDPVal}{< 0.01\xspace}

\newcommand{\ExecPongCoverageVDPVal}{< 0.01\xspace}

\newcommand{\ExecRioShootoutCoverageVDPVal}{< 0.01\xspace}

\newcommand{\ExecSnakeCoverageVDPVal}{< 0.01\xspace}

\newcommand{\ExecBirdShooterCoverageVDNonTrivial}{\textbf{0.91\xspace}}

\newcommand{\ExecBrainGameCoverageVDNonTrivial}{\textbf{1.00\xspace}}

\newcommand{\ExecCatchTheDotsCoverageVDNonTrivial}{\textbf{0.75\xspace}}

\newcommand{\ExecCatchTheGiftsCoverageVDNonTrivial}{\textbf{0.56\xspace}}

\newcommand{\ExecCatchingApplesCoverageVDNonTrivial}{\textbf{0.67\xspace}}

\newcommand{\ExecCityDefenderCoverageVDNonTrivial}{\textbf{0.87\xspace}}

\newcommand{\ExecDessertRallyCoverageVDNonTrivial}{\textbf{0.62\xspace}}

\newcommand{\ExecDieZauberlehrlingeCoverageVDNonTrivial}{\textbf{1.00\xspace}}

\newcommand{\ExecDodgeballCoverageVDNonTrivial}{\textbf{0.64\xspace}}

\newcommand{\ExecDragonsCoverageVDNonTrivial}{\textbf{0.53\xspace}}

\newcommand{\ExecEndlessRunnerCoverageVDNonTrivial}{\textbf{0.97\xspace}}

\newcommand{\ExecFallingStarsCoverageVDNonTrivial}{\textbf{0.85\xspace}}

\newcommand{\ExecFinalFightCoverageVDNonTrivial}{\textbf{0.99\xspace}}

\newcommand{\ExecFlappyParrotCoverageVDNonTrivial}{\textbf{0.79\xspace}}

\newcommand{\ExecFroggerCoverageVDNonTrivial}{\textbf{0.78\xspace}}

\newcommand{\ExecFruitCatchingCoverageVDNonTrivial}{\textbf{0.93\xspace}}

\newcommand{\ExecHackAttackCoverageVDNonTrivial}{\textbf{0.75\xspace}}

\newcommand{\ExecLineUpCoverageVDNonTrivial}{\textbf{1.00\xspace}}

\newcommand{\ExecOceanCleanupCoverageVDNonTrivial}{\textbf{0.85\xspace}}

\newcommand{\ExecPongCoverageVDNonTrivial}{\textbf{0.79\xspace}}

\newcommand{\ExecRioShootoutCoverageVDNonTrivial}{\textbf{0.96\xspace}}

\newcommand{\ExecSnakeCoverageVDNonTrivial}{\textbf{0.79\xspace}}

\newcommand{\ExecMeanCoverageVDNonTrivial}{0.82\xspace}

\newcommand{\GeneratedMutants}{243835\xspace}

\newcommand{\MutationScore}{65.62\xspace}

\newcommand{\FalsePositiveRate}{19.70\xspace}

\newcommand{\CategoryKRMKillRate}{99.18\xspace}

\newcommand{\BirdShooterMutants}{7200\xspace}

\newcommand{\BirdShooterKillRate}{96.68\xspace}
\newcommand{\BrainGameMutants}{7800\xspace}

\newcommand{\BrainGameKillRate}{40.40\xspace}
\newcommand{\CatchTheDotsMutants}{9000\xspace}

\newcommand{\CatchTheDotsKillRate}{68.79\xspace}
\newcommand{\CatchTheGiftsMutants}{8000\xspace}

\newcommand{\CatchTheGiftsKillRate}{81.53\xspace}
\newcommand{\CatchingApplesMutants}{2500\xspace}

\newcommand{\CatchingApplesKillRate}{76.92\xspace}
\newcommand{\CityDefenderMutants}{8900\xspace}

\newcommand{\CityDefenderKillRate}{11.79\xspace}
\newcommand{\DessertRallyMutants}{14569\xspace}

\newcommand{\DessertRallyKillRate}{99.65\xspace}
\newcommand{\DieZauberlehrlingeMutants}{12397\xspace}

\newcommand{\DieZauberlehrlingeKillRate}{29.57\xspace}
\newcommand{\DodgeballMutants}{7821\xspace}

\newcommand{\DodgeballKillRate}{84.22\xspace}
\newcommand{\DragonsMutants}{23838\xspace}

\newcommand{\DragonsKillRate}{98.27\xspace}
\newcommand{\EndlessRunnerMutants}{13199\xspace}

\newcommand{\EndlessRunnerKillRate}{46.21\xspace}
\newcommand{\FallingStarsMutants}{13162\xspace}

\newcommand{\FallingStarsKillRate}{87.41\xspace}
\newcommand{\FinalFightMutants}{20650\xspace}

\newcommand{\FinalFightKillRate}{90.41\xspace}
\newcommand{\FlappyParrotMutants}{3800\xspace}

\newcommand{\FlappyParrotKillRate}{38.58\xspace}
\newcommand{\FroggerMutants}{14600\xspace}

\newcommand{\FroggerKillRate}{92.97\xspace}
\newcommand{\FruitCatchingMutants}{6279\xspace}

\newcommand{\FruitCatchingKillRate}{66.22\xspace}
\newcommand{\HackAttackMutants}{8200\xspace}

\newcommand{\HackAttackKillRate}{40.44\xspace}
\newcommand{\LineUpMutants}{5600\xspace}

\newcommand{\LineUpKillRate}{98.61\xspace}
\newcommand{\OceanCleanupMutants}{10600\xspace}

\newcommand{\OceanCleanupKillRate}{63.17\xspace}
\newcommand{\PongMutants}{1300\xspace}

\newcommand{\PongKillRate}{34.00\xspace}
\newcommand{\RioShootoutMutants}{10900\xspace}

\newcommand{\RioShootoutKillRate}{54.43\xspace}
\newcommand{\SnakeMutants}{7000\xspace}

\newcommand{\SnakeKillRate}{49.21\xspace}
\newcommand{\SnowballFightMutants}{4596\xspace}

\newcommand{\SnowballFightKillRate}{70.39\xspace}
\newcommand{\SpaceOdysseyMutants}{7500\xspace}

\newcommand{\SpaceOdysseyKillRate}{42.37\xspace}
\newcommand{\WhackAMoleMutants}{14424\xspace}

\newcommand{\WhackAMoleKillRate}{78.31\xspace}

\newcommand{\BirdShooterFalsePositivesRate}{22.00\xspace}

\newcommand{\BrainGameFalsePositivesRate}{4.00\xspace}

\newcommand{\CatchTheDotsFalsePositivesRate}{0.00\xspace}

\newcommand{\CatchTheGiftsFalsePositivesRate}{0.00\xspace}

\newcommand{\CatchingApplesFalsePositivesRate}{1.00\xspace}

\newcommand{\CityDefenderFalsePositivesRate}{0.00\xspace}

\newcommand{\DessertRallyFalsePositivesRate}{33.00\xspace}

\newcommand{\DieZauberlehrlingeFalsePositivesRate}{18.00\xspace}

\newcommand{\DodgeballFalsePositivesRate}{26.26\xspace}

\newcommand{\DragonsFalsePositivesRate}{99.00\xspace}

\newcommand{\EndlessRunnerFalsePositivesRate}{5.00\xspace}

\newcommand{\FallingStarsFalsePositivesRate}{13.13\xspace}

\newcommand{\FinalFightFalsePositivesRate}{97.00\xspace}

\newcommand{\FlappyParrotFalsePositivesRate}{0.00\xspace}

\newcommand{\FroggerFalsePositivesRate}{2.00\xspace}

\newcommand{\FruitCatchingFalsePositivesRate}{13.19\xspace}

\newcommand{\HackAttackFalsePositivesRate}{3.00\xspace}

\newcommand{\LineUpFalsePositivesRate}{10.00\xspace}

\newcommand{\OceanCleanupFalsePositivesRate}{44.00\xspace}

\newcommand{\PongFalsePositivesRate}{10.00\xspace}

\newcommand{\RioShootoutFalsePositivesRate}{16.00\xspace}

\newcommand{\SnakeFalsePositivesRate}{3.00\xspace}

\newcommand{\SnowballFightFalsePositivesRate}{5.00\xspace}

\newcommand{\SpaceOdysseyFalsePositivesRate}{26.00\xspace}

\newcommand{\WhackAMoleFalsePositivesRate}{42.00\xspace}
\newcommand{\GeneratedMutantsMean}{9753.40\xspace}

\newcommand{\FullCoverageBlowOneHourNEAT}{5\xspace}
\newcommand{\WinningStatesBirdShooterNEAT}{30\xspace}
\newcommand{\WinningStatesBirdShooterRandom}{1\xspace}
\newcommand{\WinningStatesBrainGameNEAT}{30\xspace}
\newcommand{\WinningStatesBrainGameRandom}{3\xspace}
\newcommand{\WinningStatesCatchingApplesNEAT}{30\xspace}
\newcommand{\WinningStatesCatchingApplesRandom}{19\xspace}
\newcommand{\WinningStatesCatchTheDotsNEAT}{30\xspace}
\newcommand{\WinningStatesCatchTheDotsRandom}{15\xspace}
\newcommand{\WinningStatesCatchTheGiftsNEAT}{30\xspace}
\newcommand{\WinningStatesCatchTheGiftsRandom}{20\xspace}
\newcommand{\WinningStatesCityDefenderNEAT}{3\xspace}
\newcommand{\WinningStatesCityDefenderRandom}{0\xspace}
\newcommand{\WinningStatesDessertRallyNEAT}{0\xspace}
\newcommand{\WinningStatesDessertRallyRandom}{0\xspace}
\newcommand{\WinningStatesDieZauberlehrlingeNEAT}{28\xspace}
\newcommand{\WinningStatesDieZauberlehrlingeRandom}{0\xspace}
\newcommand{\WinningStatesDodgeballNEAT}{0\xspace}
\newcommand{\WinningStatesDodgeballRandom}{0\xspace}
\newcommand{\WinningStatesDragonsNEAT}{12\xspace}
\newcommand{\WinningStatesDragonsRandom}{11\xspace}
\newcommand{\WinningStatesEndlessRunnerNEAT}{11\xspace}
\newcommand{\WinningStatesEndlessRunnerRandom}{5\xspace}
\newcommand{\WinningStatesFallingStarsNEAT}{30\xspace}
\newcommand{\WinningStatesFallingStarsRandom}{0\xspace}
\newcommand{\WinningStatesFinalFightNEAT}{30\xspace}
\newcommand{\WinningStatesFinalFightRandom}{0\xspace}
\newcommand{\WinningStatesFlappyParrotNEAT}{30\xspace}
\newcommand{\WinningStatesFlappyParrotRandom}{4\xspace}
\newcommand{\WinningStatesFroggerNEAT}{15\xspace}
\newcommand{\WinningStatesFroggerRandom}{0\xspace}
\newcommand{\WinningStatesFruitCatchingNEAT}{30\xspace}
\newcommand{\WinningStatesFruitCatchingRandom}{0\xspace}
\newcommand{\WinningStatesHackAttackNEAT}{20\xspace}
\newcommand{\WinningStatesHackAttackRandom}{0\xspace}
\newcommand{\WinningStatesLineUpNEAT}{30\xspace}
\newcommand{\WinningStatesLineUpRandom}{0\xspace}
\newcommand{\WinningStatesOceanCleanupNEAT}{27\xspace}
\newcommand{\WinningStatesOceanCleanupRandom}{14\xspace}
\newcommand{\WinningStatesPongNEAT}{30\xspace}
\newcommand{\WinningStatesPongRandom}{17\xspace}
\newcommand{\WinningStatesRioShootoutNEAT}{30\xspace}
\newcommand{\WinningStatesRioShootoutRandom}{30\xspace}
\newcommand{\WinningStatesSnakeNEAT}{0\xspace}
\newcommand{\WinningStatesSnakeRandom}{0\xspace}
\newcommand{\WinningStatesSnowballFightNEAT}{17\xspace}
\newcommand{\WinningStatesSnowballFightRandom}{0\xspace}
\newcommand{\WinningStatesSpaceOdysseyNEAT}{9\xspace}
\newcommand{\WinningStatesSpaceOdysseyRandom}{0\xspace}
\newcommand{\WinningStatesWhackAMoleNEAT}{19\xspace}
\newcommand{\WinningStatesWhackAMoleRandom}{0\xspace}
\newcommand{\WinningStatesAverageRandom}{5\xspace}
\newcommand{\WinningStatesAverageNEAT}{20\xspace}